\documentclass[10pt,fleqn,a4paper]{article}

\pdfoutput=1

\usepackage{amsmath}
\usepackage{amssymb}
\usepackage{graphicx}
\usepackage[numbers]{natbib}
\usepackage{titlesec}
\usepackage{amsmath}
\usepackage{dsfont}
\usepackage{amsthm}
\usepackage{multicol}
\usepackage{bold-extra}
\usepackage[title]{appendix}
\usepackage[usenames,dvipsnames,svgnames,table]{xcolor}
\usepackage[margin=30pt,font=small,labelfont=bf,labelsep=endash]{caption}

\newcommand{\etc}{{\it etc}}
\newcommand{\cPVM}{{\cP_{\scriptscriptstyle{\rm VM}}}}
\newcommand{\cPFP}{{\cP_{\scriptscriptstyle{\rm FP}}}}
\newcommand{\cFVM}{{\cF_{\scriptscriptstyle{\rm VM}}}}
\newcommand{\cFFP}{{\cF_{\scriptscriptstyle{\rm FP}}}}
\newcommand{\tauFP}{{\tau_{\scriptscriptstyle{\rm FP}}}}
\newcommand{\tauVM}{{\tau_{\scriptscriptstyle{\rm VM}}}}
\newcommand{\trans}[1]{{#1}^{\scriptscriptstyle{\rm T}}}

\numberwithin{equation}{section}

\begin{document}
\title{{{Stochastic Dynamics of the  Multi--State Voter Model over
      a Network based on Interacting Cliques and Zealot Candidates}}}
\author{
\\[0.3cm]
{{{Filippo Palombi$^{a}$\footnote{Corresponding
        author. E--mail: {\tt filippo.palombi@enea.it}}\ \  and Simona Toti$^{b}$}}}\\[2.0ex]
 {{\small{$^a$ENEA -- Italian Agency for New Technologies, Energy and}}}\\
{{\small{Sustainable Economic Development,}}}
 {\small {{\it Via Enrico Fermi 45, 00040 Frascati -- Italy}}}\\[.2cm]
 {{\small{$^b$ISTAT -- Istituto Nazionale di Statistica,}}}
 {\small {\it Via Cesare Balbo 16, 00184 Rome -- Italy}}\\[.4cm]
}

\date{April 2014}

\newcommand{\cA}{{\cal A}}
\newcommand{\cB}{{\cal B}}
\newcommand{\cC}{{\cal C}}
\newcommand{\cD}{{\cal D}}
\newcommand{\cE}{{\cal E}}
\newcommand{\cF}{{\cal F}}
\newcommand{\cG}{{\cal G}}
\newcommand{\cH}{{\cal H}}
\newcommand{\cI}{{\cal I}}
\newcommand{\cJ}{{\cal J}}
\newcommand{\cK}{{\cal K}}
\newcommand{\cL}{{\cal L}}
\newcommand{\cM}{{\cal M}}
\newcommand{\cN}{{\cal N}}
\newcommand{\cO}{{\cal O}}
\newcommand{\cP}{{\cal P}}
\newcommand{\cQ}{{\cal Q}}
\newcommand{\cR}{{\cal R}}
\newcommand{\cS}{{\cal S}}
\newcommand{\cT}{{\cal T}}
\newcommand{\cU}{{\cal U}}
\newcommand{\cV}{{\cal V}}
\newcommand{\cW}{{\cal W}}
\newcommand{\cX}{{\cal X}}
\newcommand{\cY}{{\cal Y}}
\newcommand{\cZ}{{\cal Z}}
\newcommand{\indicator}[1]{{\mathds{I}\{#1\}}}
\newcommand{\dE}{\mathds{E}}
\newcommand{\dP}{\mathds{P}}
\newcommand{\dR}{\mathds{R}}
\newcommand{\dN}{\mathds{N}}
\newcommand{\rd}{\text{d}}
\newcommand{\re}{\text{e}}
\newcommand{\ri}{\text{i}}
\newcommand{\rO}{\text{O}}
\newcommand{\rr}{\text{r}}
\newcommand{\rx}{\text{x}}
\newcommand{\eff}{\text{eff}}
\newcommand{\diag}{\text{diag}}
\newcommand{\off}{\text{off--diag}}
\newcommand{\ml}{m\phantom{l}}

\newcommand{\ie}{{\it i.e.\ }}
\newcommand{\eg}{{\it e.g.\ }}
\newcommand{\cfr}{{\it cf.\ }}
\newcommand{\adj}[1]{\text{adj}_V(#1)}
\newcommand{\gen}{{\Omega_{\text{vt}}}}
\newcommand{\St}{{S_{\text{vt}}}}
\newcommand{\genp}{{\Omega}}
\newcommand{\Erdos}{{Erd\H{o}s}}
\newcommand{\dPer}{\mathds{P}_{\rm er}}
\newcommand{\E}{\mathds{E}}

\maketitle
\begin{abstract}
The stochastic dynamics of the multi--state voter model is
investigated on a class of complex networks made of
non--overlapping cliques, each hosting a political candidate and
interacting with the others via \Erdos--R\'enyi links. Numerical
simulations of the model are interpreted in terms of an {\it
  ad--hoc} mean field theory, specifically tuned to resolve the
inter/intra--clique interactions. Under a proper definition of the
thermodynamic limit (with the average degree of the agents kept
fixed while increasing the network size), the model is found to
display the empirical scaling discovered by Fortunato and
Castellano (2007) \cite{fcscaling}, while the vote distribution resembles
roughly that observed in Brazilian elections.   
\end{abstract}

\section{Introduction}

In a fairly recent paper \cite{fcscaling}, Fortunato and Castellano (FC) show
that the excess of the number of votes received in proportional elections by
candidates over the average number of votes per intra--party competitor has a
universal distribution, invariant in time and  independent of the country.
The analysis performed in~\cite{fcscaling} focuses in particular on the
political elections held along the past fifty years in Italy, Poland and
Finland. In order to interpret the empirical evidence, FC propose a {\it word
  of mouth} model, where political opinions originate from the candidates and
propagate down stochastic trees. With an appropriate choice of the  underlying
parameters, the model reproduces faithfully the observed voting behavior.   

A more recent extension of the above analysis to a larger pool of countries
\cite{fempanal} confirms the universal trend in Denmark and Estonia, but
proves it to be hopelessly broken in countries such as Brazil, Slovenia,
Greece, \etc. In each of these, the probability law of the intra--party excess
of votes displays specific deviations from the universal FC curve,
while persisting in time. Therefore, the discrepancies --- as the authors
of~\cite{fempanal}  claim~--- are {\sl ``likely to be due to peculiar
  differences in the election rules''}. An interesting and 
not understood case is indeed represented by Brazil, which has been
first studied in \cite{CostaFilho}. Here, voting is compulsory,
though a justification form for not voting can be filled at
election centers and post offices. Compulsory voting can be hardly
encoded in the word of mouth model by FC without introducing major
modifications, thus it makes sense to approach the Brazilian case
differently. The search for a dynamical model reproducing at least
qualitatively the Brazilian pattern  is the main phenomenological
motivation for the  present paper.  

Notoriously, a model where voting is compulsory is the voter model
(VM)~\cite{Clifford1973,Holley1975}, in that here every agent expresses a 
preference at each time step of the stochastic dynamics. In
addition, political positions in the VM differ only by their tag, just like
candidates in the word of mouth model. The reason why a simple toy model such
as the VM could effectively reproduce the universalities of real elections 
and their breaking patterns lies in the guiding principle of complexity
theory that a {\it complex} behavior often emerges from {\it simple} rules. Of course, the 
model needs to be properly adapted to catch the main features of the intra--party
dynamics in proportional elections with compulsory voting, yet much is already 
present in the literature:
\begin{itemize}
\item{First of all, the VM is a binary model, unfit to describe
    elections with more than two candidates. This difficulty is
    easily overcome by turning to the multi--state variant
    considered with different purposes in
    \cite{Bohme,Hubbell,McKane,Pigolotti}.} 
\item{Secondly, the stochastic dynamics of both the binary VM and
    its multi--state variant drives ultimately the system towards
    consensus, a uniform state where all agents share the same
    political preference. A quantitative description of consensus
    formation in the multi--state VM has been recently given
    in~\cite{Starnini} in terms of mean field theory
    (MFT). Needless to say, consensus states have been nowhere
    near observed in political elections, thus they should not be
    contemplated by any serious attempt to describe empirical data
    in a semi--realistic fashion. The simplest way to avoid them
    seems to be the use of stubborn agents (also known as {\it
      zealots}), that is to say agents who never change political
    preference, while taking part in the stochastic dynamics as
    opinion donors. Zealots have been originally introduced in
    \cite{Mobilia2003} in the framework of the binary VM and have
    been subsequently used to investigate several aspects of
    opinion dynamics (see for instance
    \cite{Acemoglu2010,Yildiz2012,Wu2012,Xie1,Xie2,Singh}). 
    They have also been used in the context of the multi--state VM, see for instance  
    \cite{mobilia2013} for an analysis of the three--party constrained VM in presence
    of committed agents. In principle zealots can model both political activists and 
    candidates. In the present paper, we use them as list candidates.} 
\item{Thirdly, if candidates exert their influence on the
    neighbor voters by letting them change opinion according to
    the passive interaction rule defining the VM, their effect at
    macroscopic level depends certainly on the topological
    structure of the social network they are part of and the
    long--range competition they engage with other
    candidates. This is to a large extent unexplored to--date. We
    consider a voter network partitioned into non--overlapping
    cliques (sub--networks where the agents are all pairwise
    connected), all having the same size. Each clique hosts one
    candidate and candidates belong all to the same party. Voters
    belonging to different cliques are then stochastically
    connected via \Erdos--R\'enyi links. This kind of network
    resembles somewhat the small-worlds model by Watts--Strogatz
    \cite{wattz1998}, the differences being that {\it i}) cliques
  in the latter are overlapping before rewiring, {\it ii})~in our model there
  is no {\it re}--wiring at all, {\it iii})  the Watts--Strogatz
  model converges to an \Erdos--R\'enyi network as the rewiring
  probability increases, whereas our network model converges to an
  almost complete graph upon increasing the frequency of the
  inter--clique links. In a sense, cliques play in our proposal a
  r\^ole analogous to the stochastic trees in the word of mouth
  model by FC. Yet, voters belonging to different trees never
  interact directly, whereas inter--clique interactions are
  essential to the opinion dynamics of our model.} 
\end{itemize}

We study the distribution of the excess of votes at equilibrium by
means of Monte Carlo simulations and, in a range of values of the
model parameters, we observe distributions resembling
roughly that of Brazilian elections. An analytic description
of the numerical results is at least partly possible thanks to
MFT, but in order to take into consideration the specific topology
of the network, MFT has to be formulated in terms of clique
variables. Using the standard machinery of the Master
Equation~\cite{Gardiner}, we perform a Kramers--Moyal expansion
and derive the corresponding Fokker--Planck (FP) equation. The
drift equations obtained by suppressing the diffusion term in the
latter can be analytically solved. Average values are thus
calculated exactly and show a perfect agreement with those
obtained numerically. In order to solve the complete FP equation,
we simulate the stochastic process associated to it and discuss
the effect of time discretization by considering two different
recipes to enforce boundary reflections. 

Here is a plan of the paper. In sect.~2, we define precisely the network, the stochastic dynamics 
and the thermodynamic limit of the model. In sect.~3, we report on some Monte Carlo simulations of
the autocorrelation time, the average values of the clique variables and the distribution
of the intra--party excess of votes. In sect.~4, we derive the FP equation and calculate the average values in MFT. 
Sect.~5 is devoted to a discussion of the stochastic process associated to the FP equation. Conclusions
are drawn in sect.~6. In appendix A, we discuss some theoretical aspects of an expansion of
the probability density of MFT based on Dirichlet
distributions. Specifically, we examine the analytic structure of the
contributions to the distribution of the excess of votes arising from it. 

\section{Definition of the model}

\subsection{Voter network}

We consider a social network described by an undirected graph $\cG =
(V,E)$. An element $x\in V$ represents a voter, while an edge $(x,y)\in E$
indicates the acquaintance $x,y\in V$ share. Each voter expresses a
political preference as an element of $\cW = \{1,2,\ldots,Q\}$, with $Q\ge 2$. We
denote by $X = \cW^V$  the space of voter configurations: if $\eta(\cdot)\in X$ and
$x\in V$, then $\eta(x)$ represents the political preference of $x$ in
$\eta(\cdot)$. We assume that $Z = \{z_1,\ldots,z_Q\}\subset V$ is a special
subset of voters. Elements of $Z$ are identified with list candidates and fulfill $\eta(z_i)=i$ at all
times, \ie the state space is actually a proper subset $\bar X\subset X$.
 We denote by $V_0 = V\setminus Z$ the set of the {\it dynamic} voters. We also assume that $\cG$ 
decomposes into $Q$ cliques of equal size, each hosting a candidate. More
precisely, we first introduce a partition $\{C_i\}_{i=1}^Q$,  with $C_i\subset
V$ and
\begin{align}
& \bullet \quad  V = \bigcup_{i=1}^Q C_i\,;\\
& \bullet \quad  C_i\cap C_j = \emptyset \text{\ \ and\ \ } |C_i|=|C_j| \text{\ \ if \ \ } i\ne j\,;\\[2.0ex]
& \bullet \quad  z_i\in C_i\,, \quad i=1,\ldots,Q\,.
\end{align}
Then, we group edges according to the position of their end--points in $\{C_i\}$ by defining the sets
\begin{equation}
E_{ij} = \{(x,y)\in E: \ \ x\in C_i \text{\ and } y\in C_j\}\,.
\end{equation}
We finally request $(x,y)\in E_{ii}$,  $\forall x,y\in
C_i$. In addition, we assume that candidates are disconnected from cliques different than 
their own: given $1\le i,j\le Q$ and $i\ne j$, $(z_i,y)\notin E_{i,j}$,
$\forall y\in C_j$. Connections between different cliques are random and modeled
according to an \Erdos--R\'enyi model with link probability $p\in(0,1]$, namely
\begin{equation}
\text{given \ } i\ne j\,,\ x\in C_i/\{z_i\} \text{\ \ and \ } y\in
C_j/\{z_j\}\,,\text{\ \ then \ } \text{prob}\{(x,y)\in E_{ij}\} = p\,.
\end{equation}
We now introduce some definitions, which set--up the language of the clique MFT and are used in the sequel. 
Given $\eta(\cdot)\in \bar X$, we define 
\begin{equation}
\eta_k = \{x\in V:\ \ \eta(x) = k\}\,,\qquad k=1,\ldots,Q\,.
\end{equation}
Accordingly, $v_k=|\eta_k|$ represents the number of votes for
the candidate $z_k$ in $\eta(\cdot)$. Analogously, we define
\begin{equation}
\eta_k^{(i)} = \{x\in C_i:\ \ \eta(x) = k\}\,,\qquad i,k=1,\ldots,Q\,.
\end{equation}
Again, $\eta_k^{(i)}$ represents the subset of voters for the candidate $z_k$
belonging to $C_i$ and we denote by $v_k^{(i)} = |\eta_k^{(i)}|$ the number of
votes for $z_k$ within $C_i$. From our definitions, it follows trivially that
$\eta_k=\cup_{i=1}^Q \eta_k^{(i)}$ and $V = \cup_{k=1}^Q\eta_k$. Note
that $1\le v_k^{(k)} \le |V|/Q$ since $z_k\in C_k$, whereas $0\le v_k^{(i)}\le |V|/Q-1$ for $k\ne i$
owing to the presence to the opposing zealot $z_i$. By consistency, we have
\begin{equation}
v_k = \sum_{i=1}^Q v_k^{(i)}\,.
\end{equation}
Along the same lines, we define
\begin{equation}
\eta_{km} = \{(x,y)\in E:\ \ \eta(x) = k \text{\ \ and \ }
\eta(y)=m\}\,,\qquad k,m=1,\ldots,Q.
\end{equation}
For $k\ne m$, $\eta_{km}$ represents the subset of edges for
which a voter for $z_k$ may turn into a voter for $z_m$ in
$\eta(\cdot)$ and the other way round. Analogously, we define
\begin{equation}
\eta_{km}^{(ij)} = \{(x,y)\in E_{ij}:\ \ \eta(x) = k \text{\ \ and \ }
\eta(y)=m\}\,,\qquad i,j,k,m=1,\ldots,Q,
\end{equation}
and for $k\ne m$, we interpret $\eta_{km}^{(ij)}$ as before, except that
now the transition may only occur owing to an interaction between a voter in $C_i$ and
one in $C_j$. Obviously, we have $\eta_{km} =
\cup_{i,j=1}^Q \eta_{km}^{(ij)}$. 

\begin{figure}[t!]
  \begin{minipage}[!t]{0.48\textwidth}
    \centering
    \includegraphics[width=0.85\textwidth]{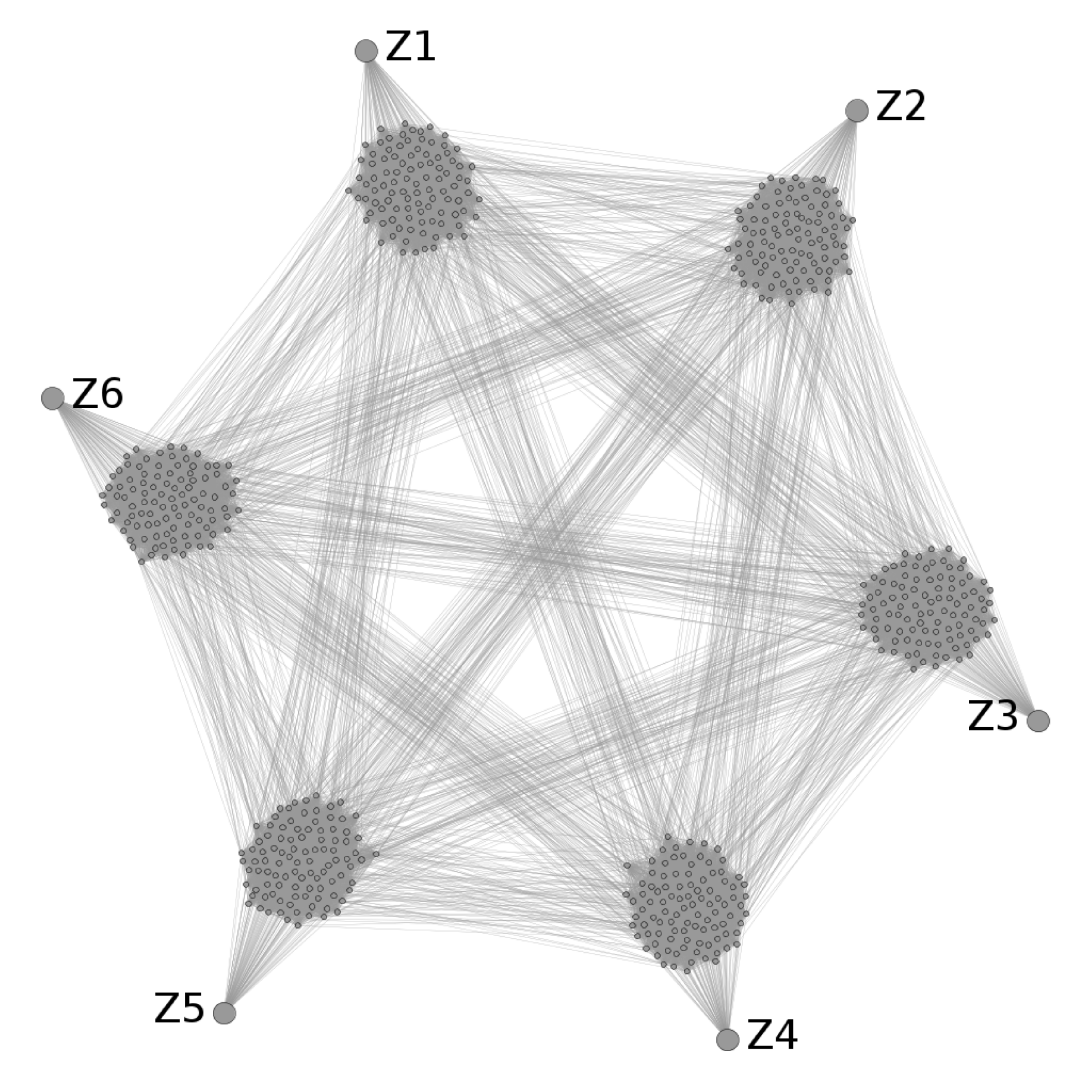}
  \end{minipage}
  \begin{minipage}[!t]{0.48\textwidth}
    \centering
    \includegraphics[width=0.65\textwidth]{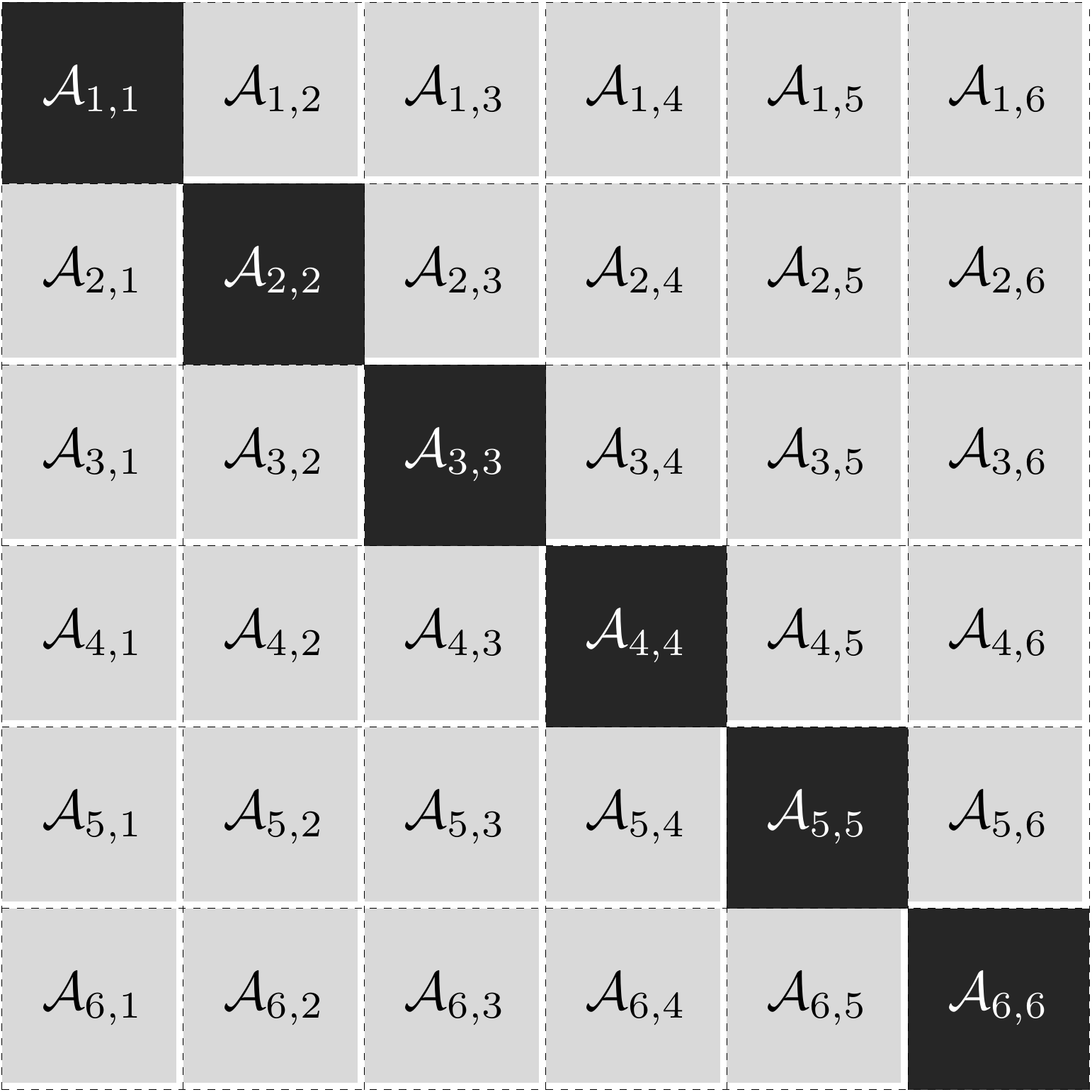}
  \end{minipage}
  \vskip 0.3cm
  \centering
  \hskip -0.8cm\includegraphics[width=0.8\textwidth]{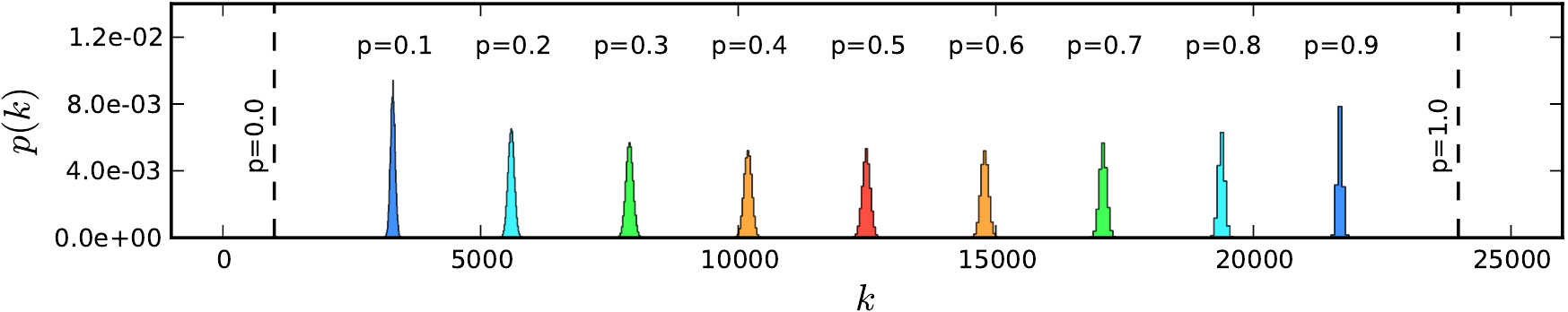}
  \vskip -0.0cm
  \caption{ \footnotesize {\it Top left}: an illustrative network sample with $|V|=600$, $Q=6$ and $p=0.01$.
    Candidates are highlighted. {\it Top right}: average structure of the adjacency matrix $\cA$ of the network at $Q=6$
    and ordering function $s$ as in eqs.~(\ref{eq:ordone})--(\ref{eq:ordtwo}); black (gray) squares describe dense intra--clique (sparse inter--clique) links. 
    {\it Bottom}: distributions of the agent degree $k$ for a network with $|V|=24,\!000$ and $Q=24$.}
\end{figure}

\vskip 0.1cm

In Fig.~1 (top left) we show a network sample with $|V|=600$, $Q=6$ and
$p=0.01$\footnote{the graph has been produced by using {\tt Gephi} with a Force Atlas layout, see ref.~\cite{ICWSM09154}},
where candidates are highlighted. In the same figure (top right), we show the
average structure of the adjacency matrix $\cA$ of the network model at $Q=6$, corresponding 
to sorting the elements of $V$ according to an ordering function
$s:V\to\{0,1,\ldots,|V|-1\}$, such that 
\begin{align}
& \bullet \ (k-1)\frac{|V|}{Q}\le
    s(x)<k\frac{|V|}{Q}\,,\quad \text{if \ \ } x\in C_k\,,\qquad k=1,\ldots,Q\,;\label{eq:ordone}\\
& \bullet \ s(z_k) = (k-1)\frac{|V|}{Q}\,,\qquad k=1,\ldots,Q\,.
\label{eq:ordtwo}
\end{align}
Diagonal blocks $\cA_{kk}$ represent intra--clique acquaintances, probabilistic \Erdos--R\'enyi
links populate the off--diagonal blocks $\cA_{ik}$ and thick--white borders indicate the
absence of extra--clique zealot--voter and extra--clique zealot--zealot
links. Again in Fig.~1 (bottom), 
we show some distributions of the voter degree for a network with $|V|=24,\!000$ and $Q=24$.

As mentioned in the introduction, this network model looks somewhat similar to the 
construction by Watts--Strogatz, provided we interpret $p$ as the analogous of the 
rewiring probability, yet the reader will notice some relevant differences:
\begin{itemize}
\item[$i$)]{both our network and the Watts--Strogatz one have a
    regular structure at $p=0$, with the agents grouped into
    cliques. Nevertheless, in the Watts--Strogatz model cliques
    overlap partially, whereas in ours they are fully disconnected;}
\item[$ii$)]{as $p$ increases, the Watts--Strogatz model becomes progressively
    disordered and it eventually degenerates to a random graph at $p=1$. By
    contrast, the level of disorder of our network increases only up to
    $p=1/2$. At larger values of $p$ the network becomes progressively reordered,
    until it turns into an almost complete graph at $p=1$ (recall that $Z$ does 
    not take part in the random wiring process);}  
\item[$iii$)]{the average degree of our network is monotonic increasing in $p$.}
\end{itemize}

\subsection{Microscopic dynamics}

A multi--state voter dynamics is assumed to run on top of $\cG$. A
voter $x\in V_0$ waits an exponential random time with rate $\gamma=1$, at
which she flips to the vote of one of her neighbors, chosen with
flat probability. The microscopic state at time $t$ is an element $\eta(\cdot,t)\in \bar X$. 
As already stated in the previous section, zealots keep their vote constant,
\ie $\eta(z_i,t)=i$ for $i=1,\ldots,Q$ and $t\in [0,\infty)$. All other observables change in 
time. The microscopic dynamics is rigorously defined in terms of the Markov generator of the process,  
\begin{equation}
\genp f(\eta) =
\sum_{x\in\,V_0}\,\sum_{k\in\cW}c_\eta(x,k)\{f(\eta_{x,k})-f(\eta)\}\,,
\end{equation}
an operator acting on the set $C(\bar X)$ of the cylinder functions on $\bar X$. Here, 
we denote by
\begin{equation}
\eta_{x,k}(\cdot) = \left\{\begin{array}{ll}\eta(z) & \text{if } z\ne
x\,,\\ k & \text{if } z = x\,. \end{array}\right.
\end{equation}
the microscopic state following a transition of voter $x$ in favor of candidate $z_k$, while
\begin{equation}
c_\eta(x,k) =
\indicator{\eta(x)\ne k}\frac{1}{|\adj{x}|}\sum_{y\in\adj{x}}\indicator{\eta(y)=k}\,,
\end{equation}
represents the corresponding transition rate. Note that the sum over $x$
defining the Markov generator extends to $V_0$, while the adjacency vector
$\adj{x}$ includes all neighbors of $x$ in $V$, to reflect the active r\^ole
of the candidates as opinion donors. In the sequel, we focus on the stochastic
trajectory of the vector $v=\{v_k^{(i)}\}_{i,k=1}^Q$, which evolves in a
subset of $\{0,1,\ldots,|V|/Q\}^{Q\times Q}$ subject to the constraints 
\begin{equation}
\left\{\begin{array}{l} \sum_{k=1}^Q v_k^{(i)}(t) = |V|/Q\,,\\[2.0ex]  v_i^{(i)}(t) \ge 1\,,\end{array}\right. \qquad
i=1,\ldots,Q\quad \text{and}\quad  t\in[0,\infty)\,.
\label{eq:normalization}
\end{equation}

\subsection{Thermodynamic limit}

We wish to investigate the statistical properties of the model in the thermodynamic limit
$|V|\to\infty$. In order to ensure that this limit is taken along a trajectory of constant 
physics, the geometric set--up has to be rescaled as $|V|$ increases. To this aim, we 
look at the average degree of the voters, $\E|\adj{x}|$ with $x\in V_0$, where the expectation 
is taken over the network ensemble described in sect.~2.1. It holds
\begin{equation}
\E|\adj{x}| = \delta_\text{in}+\delta_\text{out}\,,
\label{eq:avdeg}
\end{equation}
where
\begin{equation}
\delta_\text{in} =\left(\frac{|V|}{Q}-1\right)\,,\qquad\qquad \delta_\text{out}=p(Q-1)\left(\frac{|V|}{Q}-1\right)\,,
\label{eq:deltainout}
\end{equation}
represent the intra-- and extra--clique average degree. We
keep both $\delta_\text{in}$ and $\delta_\text{out}$ constant as
$|V|\to\infty$, which is trivially achieved by imposing 
\begin{equation}
{|V|}/{Q} = \omega_1\,, \qquad p(Q-1)=\omega_2\,,
\label{eq:scalingpars}
\end{equation}
with $\omega_1\gg 1$ and $\omega_2>0$ two independent constants. Each pair
$(\omega_1,\omega_2)$ makes the model approach the thermodynamic limit 
along a specific trajectory. Universalities, if there are at all, 
should be defined as invariant properties of the model in the thermodynamic limit with 
respect to different choices of $(\omega_1,\omega_2)$. For later convenience, it is 
worthwhile defining the fractions
\begin{align}
\pi_\text{in} & =
\frac{\delta_\text{in}}{\delta_\text{in}+\delta_\text{out}} =
\frac{1}{1+p(Q-1)} = \frac{1}{1+\omega_2}\,,\\[2.0ex]
\pi_\text{out} &
=\frac{\delta_\text{out}}{\delta_\text{in}+\delta_\text{out}}=\frac{p(Q-1)}{1+p(Q-1)}
= \frac{\omega_2}{1+\omega_2}\,,
\end{align}
representing the probabilities of intra-- and inter--clique
interaction, \ie the probabilities that a change of $\eta(x)$ occurs due to an 
interaction of $x$ with a voter belonging to her same clique or to a different one. 

\subsection{Scaling variables}

In the thermodynamic limit, the vote vector $v$ has an increasingly large
number of components, each ranging between 0 and $\omega_1$  (besides
contributions from candidates). 
For this reason, we introduce the normalized variables
\begin{equation}
\phi^{(i)}_k = \frac{\ Qv^{(i)}_k}{|V|} = \omega_1^{-1}v^{(i)}_k\,,\qquad i,k=1,\ldots,Q\,.
\label{eq:normalizone}
\end{equation}
In terms of these, eq.~(\ref{eq:normalization}) reads
\begin{equation}
\left\{\begin{array}{l} \sum_{k=1}^Q \phi_k^{(i)}(t) = 1\,,\\[2.0ex]  \phi_i^{(i)}(t) \ge \omega_1^{-1}\,,\end{array}\right. \qquad
i=1,\ldots,Q\quad \text{and}\quad  t\in[0,\infty)\,.
\label{eq:normaliztwo}
\end{equation}
Thus, we see that the vector $\phi=\{\phi^{(i)}_k\}_{i,k=1}^Q$ lives on a Cartesian 
product of cut off simplices
\begin{align}
& \cT_Q(\omega_1^{-1}) = T^{(1)}_Q(\omega_1^{-1})\times\ldots\times T^{(Q)}_Q(\omega_1^{-1})\,,\\[2.0ex]
& T^{(i)}_Q(\omega_1^{-1}) = \left\{x\in \dR_+^Q : \quad  x_i\ge \omega_1^{-1}\,,\   |x|_1 = 1\right\}\,,
\end{align}
with $|x|_1 = \sum_{k=1}^Qx_k$ the taxicab norm of $x$. Owing to the above constraints, vote variables
are redundant and one can freely choose to represent one variable per clique in terms of the 
others. Symmetry suggests to eliminate the diagonal variables $\phi^{(k)}_k$. For later 
convenience, we denote by $\bar\phi = \{\phi^{(i)}_k\}_{i,k:i\ne k}^{1,\ldots, Q}$ the vector of the
essential variables following our conventional choice. Accordingly, the intra--party excess
of votes candidate $z_k$ receives is given by
\begin{equation}
\phi_k = \sum_{i=1}^Q\phi^{(i)}_k = 1 + \sum_{i\ne k}^{1\ldots Q}\phi^{(i)}_k - \sum_{i\ne k}^{1\ldots Q}\phi^{(k)}_i\,,\qquad k=1,\ldots Q\,.
\label{eq:phikFC}
\end{equation}
These are just the scaling variables considered by FC, expressed in the
language of cliques. If $\cPVM(\bar\phi)$ denotes the probability density
of the normalized vote vector at 
equilibrium, then the probability density of $\phi_k$ (which is independent of $k$ as a consequence of the symmetries of the model)
is represented by the marginal distribution
\begin{equation}
\cFVM(x) = \int_{\cT_Q(\omega_1^{-1})}\!\!\delta(\phi_k-x)\,\cPVM(\bar\phi)\,\rd\bar\phi\,\,.
\label{eq:exss}
\end{equation}
The reader should bear in mind that in the thermodynamic limit $\phi_k$ ranges over an infinite set of discrete values, 
namely $\phi_k=\omega_1^{-1}\ell$, $\ell=1,2,\ldots,|V|(1-\omega_1^{-1})$. The minimum variation of $\phi_k$ is thus 
$\Delta\phi_k=\omega_1^{-1}$. When $\Delta\phi_k\ll \phi_k$ we can legitimately consider $\phi_k$ as a continuous 
variable, but close to the lower bound $\omega_1^{-1}$ the quantization of~$\phi_k$ is unmistakable. 
Anyway, an exact representation of $\cFVM(x)$ at all scales should be given in
terms of Dirac delta functions, \ie
\begin{equation}
\cFVM(x) = \sum_{\ell=1}^\infty c_\ell\, \delta(x-\omega_1^{-1}\ell)\,,\qquad \sum_{\ell=1}^\infty c_\ell=1\,,
\end{equation}
in the thermodynamic limit.

\begin{table}[!t]
\footnotesize
  \begin{center}
    \begin{tabular}{l|l|l}
      \hline\hline\\[-1.0ex]
      $v_k^{(i)}$ & number of votes received by candidate $z_k$ from voters in the $i$--th clique & \phantom{c} \\[1.0ex]
      $\phi_k^{(i)} = \omega_1^{-1}v_k^{(i)}$ & contribution to the excess of votes of candidate $z_k$ from the $i$--th clique & eq.~(\ref{eq:normalizone}) \\[1.0ex]
      $\bar\phi = \{\phi_k^{(i)}\}_{i,k:i\ne k}^{1,\ldots,Q}$ & state vector of the system & \\[1.0ex]
      $\cPVM(\bar\phi)$ & probability density of the state vector in the VM &  \\[1.0ex]
      $\cP^{\scriptscriptstyle \text{off}}_{\scriptscriptstyle \text{VM}}(x)$ & marginal probability density of the off--diagonal state variables $\phi_k^{(i)}$, $i\ne k$ &  eq.~(\ref{eq:Poff}) \\[1.0ex]
      $\cP^{\scriptscriptstyle \text{diag}}_{\scriptscriptstyle \text{VM}}(x)$ & marginal probability density of the diagonal state variables $\phi_k^{(k)}$ &  eq.~(\ref{eq:Pdiag}) \\[1.0ex]
      $\cFVM(x)$ & marginal probability density of the excess of votes in the VM &  eq.~(\ref{eq:exss}) \\[1.0ex]
      $\cPFP(\bar\phi)$ & probability density of the state vector from the FP equation & eq.~(\ref{eq:fpeq}) \\[1.0ex]
      $\cFFP(x)$ & marginal probability density of the excess of votes from the FP equation & \\[1.0ex]
      \hline\hline
    \end{tabular}
    \caption{\footnotesize Synoptic table of the most relevant quantities considered in the present paper}
    \label{tab:summary}
  \end{center}
\end{table}

\vskip 0.5cm 

A summary of the most relevant quantities considered in the present paper is reported in Table~\ref{tab:summary}.

\section{Monte Carlo simulations of the VM}

As well known since \cite{Mobilia2007}, competing candidates prevent the 
system from reaching exit configurations. The extremal states
$\phi_k=\omega_1^{-1}$ behave like reflecting boundaries in $\bar
X$. They make the microscopic dynamics of the model run
endlessly. After a transient depending on the initial state, the
voter configuration reaches a dynamic equilibrium governed by
$\cPVM(\bar\phi)$. We explore this numerically by means of Monte
Carlo simulations.  

In the thermodynamic limit, the parameter space of the model is represented by 
the pairs $(\omega_1,\omega_2)$. At finite $|V|$ the
adjacency matrix becomes denser as $\omega_1$ and/or $\omega_2$
increase. Simulations become in this limit more and more demanding, owing to
both an increasingly severe memory requirement to host the model on a machine
and an intrinsic enhancement of the autocorrelation time, as we discuss in a
moment. According to such restrictions and the computing power in our
availability, we choose to simulate the model at
$\omega_1=1,\!000.0,2,\!000.0$ and $\omega_2=0.3,0.6,0.9$.  For each
pair  $(\omega_1,\omega_2)$, we perform simulations at five values of $|V|$,
so as to be able to observe clearly how the system approaches the
thermodynamic limit. Simulation parameters are reported in
Table~\ref{tab:pars}. For each set of values, we extract randomly 20 network
samples at $\omega_1=1,\!000.0$ and 40 network samples at $\omega_1=2,\!000.0$ and
average results over them. All simulations start from a random configuration
of voters, with political preferences assigned independently with
a flat probability of $1/Q$. 

\begin{table}[!t]
\begin{minipage}[!t]{0.5\textwidth}
\small
\raggedleft
  \begin{tabular}{cc|crc}
    \hline\hline\\[-2.0ex]
    $\omega_1$ & $\omega_2$ & $|V|$ & $Q$ & $p$ \\[0.2ex] 
    \hline\\[-2.3ex]
    1,000.0 &  0.3 &  12,000 & 12 & $2.727273\times 10^{-2}$ \\
              &      &  24,000 & 24 & $1.304348\times 10^{-2}$ \\
              &      &  48,000 & 48 & $6.382979\times 10^{-3}$ \\
              &      &  72,000 & 72 & $4.225352\times 10^{-3}$ \\
              &      &  96,000 & 96 & $3.157895\times 10^{-3}$ \\
    \hline\\[-2.3ex]
    1,000.0 &  0.6 &  12,000 & 12 & $5.454545\times 10^{-2}$ \\
              &      &  24,000 & 24 & $2.608696\times 10^{-2}$ \\
              &      &  48,000 & 48 & $1.276596\times 10^{-2}$ \\
              &      &  72,000 & 72 & $8.450704\times 10^{-3}$ \\
              &      &  96,000 & 96 & $6.315789\times 10^{-3}$ \\
    \hline\\[-2.3ex]
    1,000.0 &  0.9 &  12,000 & 12 & $8.181818\times 10^{-2}$ \\
              &      &  24,000 & 24 & $3.913043\times 10^{-2}$ \\
              &      &  48,000 & 48 & $1.914894\times 10^{-2}$ \\
              &      &  72,000 & 72 & $1.267606\times 10^{-2}$ \\
              &      &  96,000 & 96 & $9.473684\times 10^{-3}$ \\
    \hline\hline
  \end{tabular}
\end{minipage}
\begin{minipage}[!t]{0.5\textwidth}
\small
\raggedright
  \begin{tabular}{cc|crc}
    \hline\hline\\[-2.0ex]
    $\omega_1$ & $\omega_2$ & $|V|$ & $Q$ & $p$ \\[0.2ex]
    \hline\\[-2.3ex]
    2,000.0 &  0.3 &  12,000 & 6  & $6.000000\times 10^{-2}$ \\
              &      &  24,000 & 12 & $2.727273\times 10^{-2}$ \\
              &      &  48,000 & 24 & $1.304348\times 10^{-2}$ \\
              &      &  72,000 & 36 & $8.571429\times 10^{-3}$ \\
              &      &  96,000 & 48 & $6.382979\times 10^{-3}$ \\
       \hline\\[-2.3ex]
    2,000.0 &  0.6 &  12,000 & 6  & $1.200000\times 10^{-1}$ \\
              &      &  24,000 & 12 & $5.454545\times 10^{-2}$ \\
              &      &  48,000 & 24 & $2.608696\times 10^{-2}$ \\
              &      &  72,000 & 36 & $1.714286\times 10^{-2}$ \\
              &      &  96,000 & 48 & $1.276596\times 10^{-2}$ \\
       \hline\\[-2.3ex]
    2,000.0 &  0.9 &  12,000 & 6  & $1.800000\times 10^{-1}$ \\
              &      &  24,000 & 12 & $8.181818\times 10^{-2}$ \\
              &      &  48,000 & 24 & $3.913043\times 10^{-2}$ \\
              &      &  72,000 & 36 & $2.571429\times 10^{-2}$ \\
              &      &  96,000 & 48 & $1.914894\times 10^{-2}$ \\
       \hline\hline
  \end{tabular}
  \vskip 0.4cm
\end{minipage}
  \caption{\footnotesize Parameters for the simulation of the
    autocorrelation time, the average values of $\phi^{(i)}_k$ and
    the distribution $\cFVM(x)$ of the excess of votes. 
\label{tab:pars}}
\end{table}

\subsection{Autocorrelation time}

\begin{figure}[t!]
  \vskip 3.0cm
  \centering
  \includegraphics[width=0.7\textwidth]{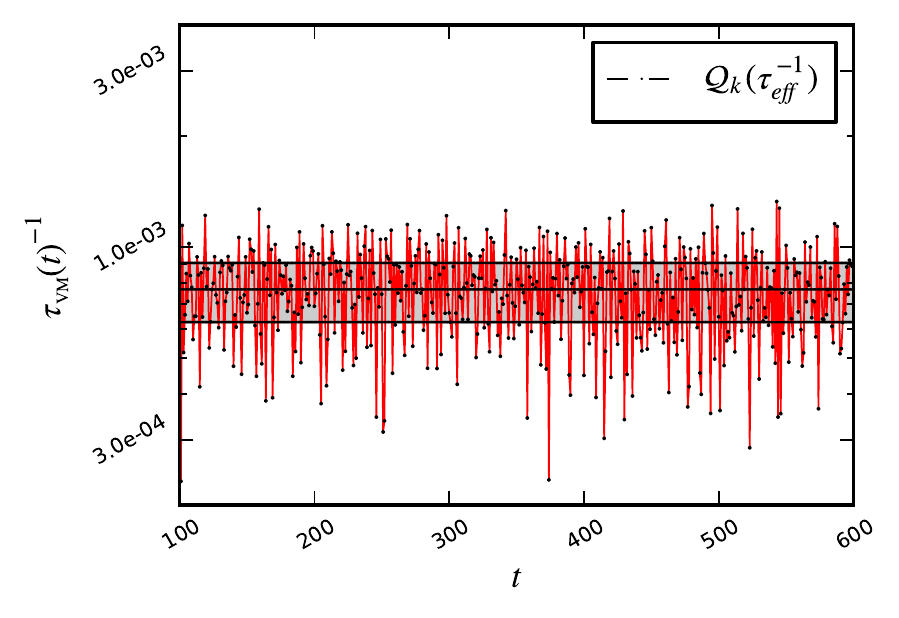}
  \vskip 1.0cm
  \hskip -0.2cm
  \includegraphics[width=0.7\textwidth]{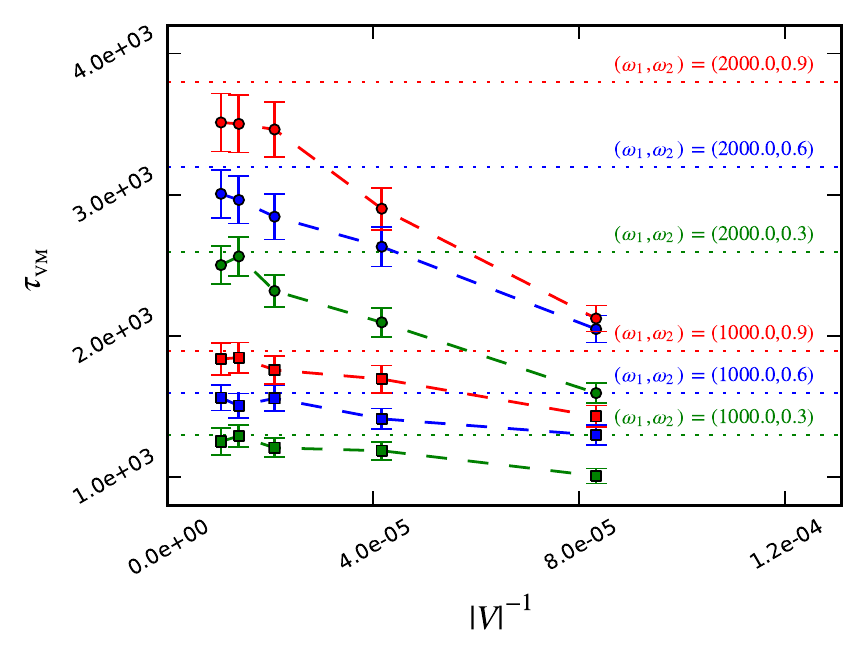}
  \caption{ \footnotesize {\it Top}: fluctuations of $\tau_\eff(t)^{-1}$
    along a single Markov chain at
    $(|V|,Q,p)=(48,\!000,48,6.383\times 10^{-3})$. {\it Bottom}: thermodynamic
    limit of $\tauVM$ for several values of
    $(\omega_1,\omega_2)$. Horizontal dotted lines correspond to
    MFT predictions. Finite size effects are compatible with ${\rm
      O}(|V|^{-1})$--terms.} 
  \label{fig:tauplateau} 
\end{figure}

As usual in the language of Monte Carlo simulations, we define a sweep as a
number $|V|-Q$ of single--voter updates, so that on average each dynamic voter
is updated once in a sweep. The autocorrelation time $\tauVM$ measures the
average number of sweeps separating two statistically independent voter
configurations. In order to simulate $\tauVM$, we proceed as follows.  Given two
generic configurations $\eta(\cdot)$ and $\bar\eta(\cdot)$, we define
their overlap $\langle\eta(\cdot)|\bar\eta(\cdot)\rangle$ as the
fraction of voters sharing the same political preference, namely   
\begin{equation}
\langle\eta(\cdot)|\bar\eta(\cdot)\rangle =
\frac{1}{|V|}\sum_{x\in V}\delta_{\eta(x),\bar\eta(x)}\,. 
\end{equation}
Hence, we define the autocorrelation function as
\begin{equation}
\cC(t) = \langle\eta(\cdot,s)|\eta(\cdot,s+t)\rangle\,.
\end{equation}
At equilibrium, $\cC(t)$ depends upon $s$ only via statistical
fluctuations, which are suppressed  by averaging over Markov
chains. It should be observed that at finite $|V|$, $\cC(t)$ does
not vanish as $t\to\infty$. Indeed, any dynamic voter has a
probability $1/Q$ of expressing the same political preference at
times $s$ and $s+t$ even though $\eta(\cdot,s)$ and
$\eta(\cdot,s+t)$ are statistically independent. Since the number
of dynamic voters is $|V|-Q$, such finite  probability gives a
contribution of $(|V|-Q)/|V|Q = (1-\omega_1^{-1})/Q$ to $\cC(t)$
as $t\to\infty$. An  additional contribution of
$Q/|V|=\omega_1^{-1}$ is due to the zealotry of the
candidates. Thus, we conclude that 
\begin{equation}
\cC(t)\ \mathop{\sim}_{t\to\infty}\ \cC_0 + \cA\,
\exp\left(-\frac{t}{\tauVM}\right)\,,\qquad \cC_0 = \frac{1}{|V|}\left[Q +
  \frac{|V|-Q}{Q}\right] = \omega_1^{-1}+\frac{1-\omega_1^{-1}}{Q}\,,
\label{eq:taueq}
\end{equation}
with an appropriate factor $\cA$, depending on the details of system but
not on $t$. In particular, we note that $\cC_0\to\omega_1^{-1}$ in
the thermodynamic limit. In order to extract $\tauVM$ from
$\cC(t)$, we introduce the effective autocorrelation time
\begin{equation}
\tau_\eff(t) = 1/\log\left\{\frac{\cC(t)-\cC_0}{\cC(t+1)-\cC_0}\right\}\,.
\label{eq:tauFP}
\end{equation}
From eq.~(\ref{eq:taueq}) it follows that $\tau_\eff(t)$ depends upon
$t$ only via fluctuations. Instead of averaging over several Monte
Carlo histories, we extract $\tauVM$ from a single Markov
chain. An example is shown in Fig.~\ref{fig:tauplateau}
(top). This plot refers to parameters
$(|V|,Q,p)=(48,\!000,48,6.383\times 10^{-3})$, corresponding to
$(\omega_1,\omega_2)=(1,\!000.0,0.3)$. It has been obtained according
to the following procedure. Starting from a random voter
configuration, we let the system relax to the equilibrium: 15,000
sweeps are largely sufficient to this aim. We then reconstruct
$\cC(t)$ from the subsequent 35,000 configurations by averaging the
overlap of all those, which are separated by a time distance of
$t$. From $\cC(t)$ we compute $\tau_\eff(t)$ and look at  the
quartiles $\cQ_k$~($k=1,2,3$) of the distribution it generates in
the range $t\in[100,600]$: larger values of $t$ can be certainly
considered, yet the statistical noise becomes increasingly
important.  We take $\tauVM = \cQ_2$ as our best estimate and
$\Delta\tauVM=\max\{\cQ_3-\cQ_2,\cQ_2-\cQ_1\}$ as its
uncertainty. Numbers depend on the network sample, therefore we
average over several networks as explained above. Standard error
propagation is applied upon averaging. 

Results for $\tauVM$ are shown in Fig.~\ref{fig:tauplateau}
(bottom) together with the MFT prediction
\begin{equation}
\tau_{\scriptscriptstyle\text{FP}} = \omega_1(1+\omega_2)(1-\omega_1^{-1})\,,
\end{equation}
which is derived in sect.~4. They admit a straightforward interpretation.
\begin{itemize}
\item{As $\omega_2\to 0$ the long--range competition among candidates reduces
    and the cliques converge to full consensus, each towards its
    own candidate. The size of the consensus domains is $\omega_1$
    close to this limit, thus one should expect
    $\tauVM\simeq\omega_1$. MFT predicts indeed $\tauFP \to
    \omega_1-1$. Candidates are responsible for the small correction of $-1$.} 
\item{Upon increasing $\omega_1$, the intra--clique consensus domains enlarge
    independent of $\omega_2$, thus making $\tauVM$ increase correspondingly.} 
\item{The long--range competition among candidates intensifies at larger
    $\omega_2$. The probability of producing consensus domains extending over
    different cliques increases correspondingly, thus enhancing $\tauVM$.} 
\item{The statistical uncertainty on $\tauVM$ increases at larger
    $\omega_2$. Since inter--clique interactions intensify as
    $\omega_2$ increases and consensus domains are progressively
    free to spread over different cliques, the variance of the
    distribution of $\tau_\eff$ along a single Markov chain
    increases as well.} 
\end{itemize}

We finally observe that the MFT prediction $\tauFP$ coincides with
$\E|\adj{x}|$ of eq.~(\ref{eq:avdeg}). This is not surprising, given the
microscopic interaction rule of the VM: it suggests
that on average a voter is ready to change political preference just when 
all her neighbors have changed their minds. 

\subsection{Marginal distributions and average votes}

\begin{figure}[!t]
  \centering
  \includegraphics[width=0.94\textwidth]{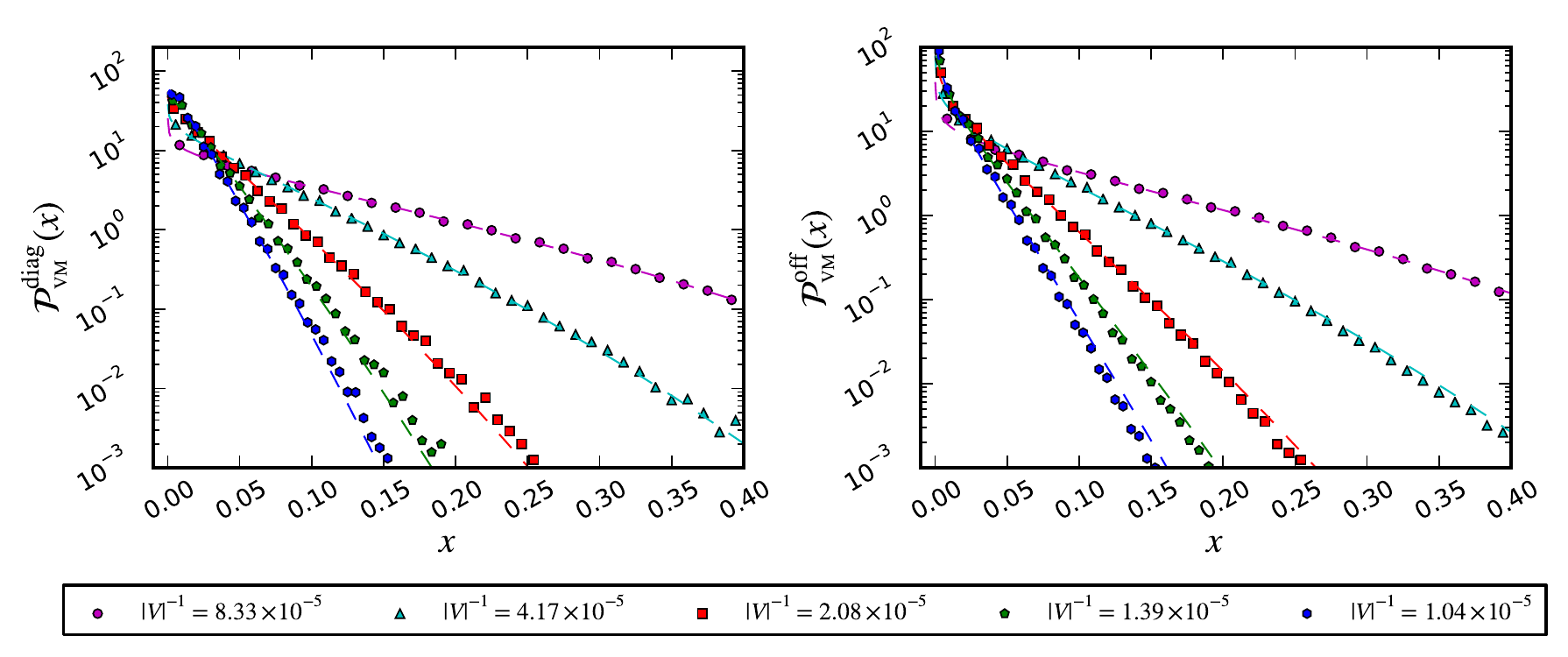}
  \vskip 0.3cm
  \centering
  \includegraphics[width=0.9\textwidth]{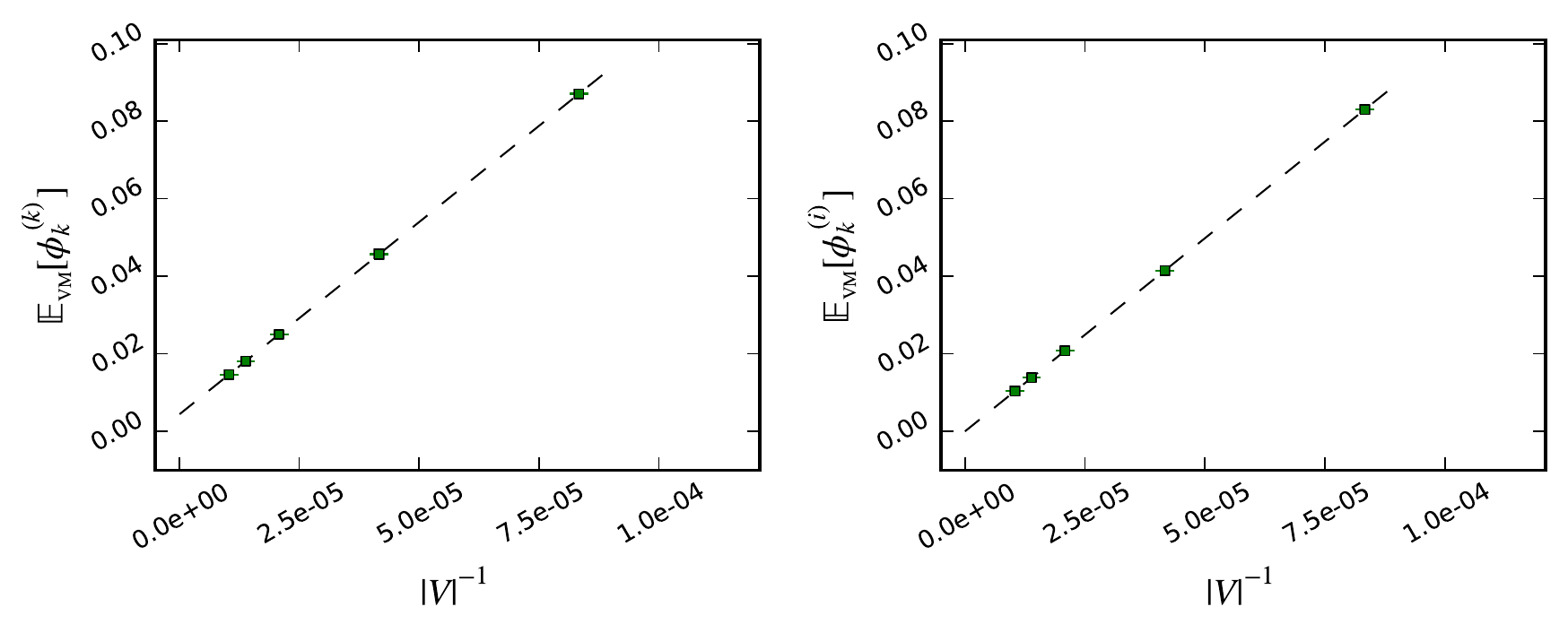}
  \vskip -0.3cm
  \caption{\footnotesize {\it Top}: marginal distribution of the
    scaling variables at
    $(\omega_1,\omega_2)=(1000.0,0.3)$. Dashed lines are empirical
    fits to Beta distributions. {\it Bottom}: average value of the
    scaling variables at
    $(\omega_1,\omega_2)=(1000.0,0.3)$. Dashed lines represent MFT
    predictions.} 
  \label{fig:marginals}  
\end{figure}

Having estimated the autocorrelation time, we can safely simulate
the equilibrium distribution of the~VM. Following an initial
thermalization, we take a snapshot of the system every $\Delta t =
k\tauFP$ sweeps, with $k$ an integer chosen so that
$\exp\{-k\}<1/Q$. In this way we ensure that the exponential
correlation of eq.~(\ref{eq:taueq}) lies below the threshold
$\cC_0$. For instance, at the largest value of $Q$ among those
reported in Table~\ref{tab:pars}, we set $k=5$. By collecting
equally large numbers of voter configurations for each network
sample at given parameters $(|V|,Q,p)$, we reconstruct the
marginal distributions of the scaling variables
\begin{align}
\label{eq:Poff}
\cP^{\scriptscriptstyle \text{off}}_{\scriptscriptstyle \text{VM}}(x) & =
\int_{\cT_Q(\omega_1^{-1})}\!\!
\delta(\phi^{(i)}_k-x)\,\cPVM(\bar\phi)\,\rd\bar\phi\,\,,\qquad  
i\ne k\,,\\[2.0ex]
\label{eq:Pdiag}
\cP^{\scriptscriptstyle \text{diag}}_{\scriptscriptstyle \text{VM}}(x) & =
\int_{\cT_Q(\omega_1^{-1})}\!\!\delta\left(1-\sum_{i\ne
    k}^{1\ldots
    Q}\phi^{(k)}_i-x\right)\,\cPVM(\bar\phi)\,\rd\bar\phi\,\,, 
\end{align}
and their expectations
\begin{align}
\E_{\scriptscriptstyle \text{VM}}[\phi^{(i)}_k] & =
  \int_{0}^{1-\omega_1^{-1}}\rd x \, x\,\cP^{\scriptscriptstyle
    \text{off}}_{\scriptscriptstyle \text{VM}}(x)\,,\qquad i\ne k\,,\\[2.0ex]
\E_{\scriptscriptstyle \text{VM}}[\phi^{(k)}_k] & =
  \int_{\omega_1^{-1}}^1\rd x \, x\,\cP^{\scriptscriptstyle
    \text{diag}}_{\scriptscriptstyle \text{VM}}(x)\,. 
\end{align}
In Fig.~\ref{fig:marginals} (top) we show $\cP^{\scriptscriptstyle
  \text{diag}}_{\scriptscriptstyle \text{VM}}$  and
$\cP^{\scriptscriptstyle \text{off}}_{\scriptscriptstyle \text{VM}}$
at $(\omega_1,\omega_2)=(1,\!000.0,0.3)$. Points represent results
of the Monte Carlo simulations, while curves correspond to
empirical fits to Beta distributions 
\begin{equation}
\beta_{a,b}(x) = \frac{\Gamma(a+b)}{\Gamma(a)\Gamma(b)}x^{a-1}(1-x)^{b-1}\,.
\label{eq:betansatz}
\end{equation}
\begin{table}[!t]
\small
\centering
  \begin{tabular}{ccrrrrr}
    \hline\hline\\[-2.0ex]
    $\omega_1$ & $\omega_2$ & \quad $Q$ \quad & $a_{\text{diag}}\hskip 0.85cm$ & $b_{\text{diag}}\hskip 0.85cm$ & $a_{\text{off}}\hskip 0.85cm$ & $b_{\text{off}}\hskip 0.85cm$ \\[0.2ex]
    \hline\\[-2.3ex]
    $1,\!000.0$ & $0.3$ & \quad 12 \quad & $8.28(3)\times 10^{-1}$ & $8.70(4)\times 10^{0}$ & $7.543(8)\times 10^{-1}$ & $8.33(1)\times 10^{0}$ \\
    & & \quad 24 \quad & $8.67(2)\times 10^{-1}$ & $1.811(5)\times 10^{1}$ & $7.178(4)\times 10^{-1}$ & $1.658(1)\times 10^{1}$ \\
    & & \quad 48 \quad & $9.45(2)\times 10^{-1}$ & $3.687(9)\times 10^{1}$ & $6.647(2)\times 10^{-1}$ & $3.138(1)\times 10^{1}$ \\
    & & \quad 72 \quad & $1.017(2)\times 10^{0}\hskip 1.5ex $ & $5.51(1)\times 10^{1}$ & $6.194(1)\times 10^{-1}$ & $4.417(1)\times 10^{1}$ \\
    & & \quad 96 \quad & $1.079(2)\times 10^{0}\hskip 1.5ex $ & $7.25(1)\times 10^{1}$ & $5.820(1)\times 10^{-1}$ & $5.553(1)\times 10^{1}$ \\
    \hline\\[-2.3ex]
    $1,\!000.0$ & $0.6$ & \quad 12 \quad & $6.74(2)\times 10^{-1}$ & $7.19(3)\times 10^{0}$ & $6.351(7)\times 10^{-1}$ & $7.005(8)\times 10^{0}$ \\
    & & \quad 24 \quad & $6.87(2)\times 10^{-1}$ & $1.488(5)\times 10^{1}$ & $6.082(4)\times 10^{-1}$ & $1.403(1)\times 10^{1}$ \\
    & & \quad 48 \quad & $7.32(2)\times 10^{-1}$ & $3.056(7)\times 10^{1}$ & $5.797(2)\times 10^{-1}$ & $2.732(1)\times 10^{1}$ \\
    & & \quad 72 \quad & $7.72(1)\times 10^{-1}$ & $4.664(9)\times 10^{1}$ & $5.580(1)\times 10^{-1}$ & $3.973(1)\times 10^{1}$ \\
    & & \quad 96 \quad & $8.31(1)\times 10^{-1}$ & $6.29(1)\times 10^{1}$ & $5.394(1)\times 10^{-1}$ & $5.138(1)\times 10^{-1}$ \\
    \hline\\[-2.3ex]
    $1,\!000.0$ & $0.9$ & \quad 12 \quad & $5.73(2)\times 10^{-1}$ & $6.16(3)\times 10^{0}$ & $5.453(6)\times 10^{-1}$ & $6.011(8)\times 10^{0}$ \\
    & & \quad 24 \quad & $5.73(2)\times 10^{-1}$ & $1.257(4)\times 10^{1}$ & $5.203(3)\times 10^{-1}$ & $1.200(1)\times 10^{1}$ \\
    & & \quad 48 \quad & $6.00(1)\times 10^{-1}$ & $2.565(7)\times 10^{1}$ & $4.970(2)\times 10^{-1}$ & $2.341(1)\times 10^{1}$ \\
    & & \quad 72 \quad & $6.46(1)\times 10^{-1}$ & $3.983(9)\times 10^{1}$ & $4.891(1)\times 10^{-1}$ & $3.481(1)\times 10^{1}$ \\
    & & \quad 96 \quad & $6.81(1)\times 10^{-1}$ & $5.38(1)\times 10^{1}$ & $4.752(1)\times 10^{-1}$ & $4.524(1)\times 10^{1}$ \\
    \hline\\[-2.3ex]
    $2,\!000.0$ & $0.3$ & \quad 6 \quad & $8.67(4)\times 10^{-1}$ & $4.29(2)\times 10^{0}$ & $8.49(2)\times 10^{-1}$ & $4.25(1)\times 10^{0}$ \\
    & & \quad 12 \quad & $8.18(5)\times 10^{-1}$ & $8.78(6)\times 10^{0}$ & $7.81(2)\times 10^{-1}$ & $8.61(2)\times 10^{0}$ \\
    & & \quad 24 \quad & $8.33(5)\times 10^{-1}$ & $1.82(1)\times 10^{1}$ & $7.58(1)\times 10^{-1}$ & $1.747(3)\times 10^{1}$ \\
    & & \quad 36 \quad & $8.43(7)\times 10^{-1}$ & $2.74(3)\times 10^{1}$ & $7.27(1)\times 10^{-1}$ & $2.550(4)\times 10^{1}$ \\
    & & \quad 48 \quad & $8.64(7)\times 10^{-1}$ & $3.68(4)\times 10^{1}$ & $7.12(1)\times 10^{-1}$ & $3.352(5)\times 10^{1}$ \\
    \hline\\[-2.3ex]
    $2,\!000.0$ & $0.6$ & \quad 6 \quad & $7.18(3)\times 10^{-1}$ & $3.57(2)\times 10^{0}$ & $7.09(1)\times 10^{-1}$ & $3.55(1)\times 10^{0}$ \\
    & & \quad 12 \quad & $6.55(5)\times 10^{-1}$ & $7.10(6)\times 10^{0}$ & $6.35(1)\times 10^{-1}$ & $6.70(2)\times 10^{0}$ \\
    & & \quad 24 \quad & $6.62(6)\times 10^{-1}$ & $1.48(1)\times 10^{1}$ & $6.21(1)\times 10^{-1}$ & $1.430(3)\times 10^{1}$ \\
    & & \quad 36 \quad & $6.78(6)\times 10^{-1}$ & $2.27(2)\times 10^{1}$ & $6.16(1)\times 10^{-1}$ & $2.159(4)\times 10^{1}$ \\
    & & \quad 48 \quad & $6.83(7)\times 10^{-1}$ & $3.02(3)\times 10^{1}$ & $6.02(1)\times 10^{-1}$ & $2.833(5)\times 10^{1}$ \\
    \hline\\[-2.3ex]
    $2,\!000.0$ & $0.9$ & \quad 6 \quad & $6.19(5)\times 10^{-1}$ & $3.08(3)\times 10^{0}$ & $6.11(2)\times 10^{-1}$ & $3.06(1)\times 10^{0}$ \\
    & & \quad 12 \quad & $5.72(5)\times 10^{-1}$ & $6.21(6)\times 10^{0}$ & $5.57(1)\times 10^{-1}$ & $6.14(2)\times 10^{0}$ \\
    & & \quad 24 \quad & $5.59(5)\times 10^{-1}$ & $1.26(1)\times 10^{1}$ & $5.32(1)\times 10^{-1}$ & $1.224(3)\times 10^{1}$ \\
    & & \quad 36 \quad & $5.65(6)\times 10^{-1}$ & $1.91(2)\times 10^{1}$ & $5.22(1)\times 10^{-1}$ & $1.829(4)\times 10^{1}$ \\
    & & \quad 48 \quad & $5.70(6)\times 10^{-1}$ & $2.55(3)\times 10^{1}$ & $5.15(1)\times 10^{-1}$ & $2.422(5)\times 10^{1}$ \\
    \hline\hline
  \end{tabular}
  \caption{\footnotesize Fit parameters of $\cP^{\scriptscriptstyle
  \text{diag}}_{\scriptscriptstyle \text{VM}}$  and
$\cP^{\scriptscriptstyle \text{off}}_{\scriptscriptstyle \text{VM}}$ vs. $\beta_{a,b}(x)$.}
\label{tab:fitpars}
\end{table}
\vskip -0.3cm
\noindent The fit constants $a$ and $b$, used in the plot and reported in
Table~\ref{tab:fitpars} for all sets of simulation parameters, have been
obtained by comparing the numerical estimates of mean and variance
with the corresponding analytic expressions known for the Beta
distribution, namely 
\begin{equation}
\E_{a,b}[x] = \frac{a}{a+b}\,,\qquad \text{var}_{a,b}[x] =
\frac{ab}{(a+b)^2(a+b+1)}\,. 
\end{equation}
The uncertainties quoted in Table~\ref{tab:fitpars} have been
obtained by bootstrapping distributions from those simulated. The
interpolation looks overall rather precise, though its quality
worsens slightly at the largest values of $|V|$. In the
thermodynamic limit the marginal distributions are indeed squeezed
towards increasingly small values of the scaling variables, where
quantization effects become non--negligible. At such low scales, any
{\it ansatz} based on a continuous density is intrinsically
unsuitable to describe the numerical data. The reader will
certainly notice that the fit parameters are far from being
uncorrelated. A close inspection to Table~\ref{tab:fitpars} 
suggests indeed the approximate distributional laws 
\begin{align}
\label{eq:betansone}
\cP^{\scriptscriptstyle \text{off}}_{\scriptscriptstyle \text{VM}}(x) & \approx
\beta_{\alpha,\bar\alpha + (Q-2)\alpha}\left(x\right)\,,\\[2.0ex]
\label{eq:betanstwo}
\cP^{\scriptscriptstyle \text{diag}}_{\scriptscriptstyle \text{VM}}(x) & \approx
\beta_{\bar\alpha,(Q-1)\alpha}\left(x\right)\,,
\end{align}
with $\alpha=\alpha(\omega_1,\omega_2,|V|)$ and $\bar\alpha =
\bar\alpha(\omega_1,\omega_2,|V|)$ being characterized by a
residual dependence on $|V|$, which is difficult to assess. We
shall come back to eqs.~(\ref{eq:betansone})--(\ref{eq:betanstwo})
in Appendix A. For the moment, we just observe that, regardless of
the analytic form of the distributions, diagonal and off--diagonal
scaling variables look very similar, though they are not
marginally identical. This makes us conclude that candidates do
not exert a very strong influence on their clique mates at the
simulation parameters considered. Nevertheless, the difference
among $\phi^{(i)}_k$ and $\phi^{(k)}_k$  increases as
$|V|\to\infty$. To check this, in Fig~\ref{fig:marginals} (bottom)
we plot the average values of the scaling  variables at
$(\omega_1,\omega_2)=(1,\!000.0,0.3)$ vs. the prediction of MFT 
\begin{equation}
\E_{\scriptscriptstyle \text{FP}}[\phi^{(i)}_k] =
\frac{\delta_{ik}\left[Q-1+\omega_2
    Q(1-\omega_1^{-1})\right]+\omega_1\omega_2(1-\omega_1^{-1})^2}{Q-1+\omega_1
  \omega_2Q(1-\omega_1^{-1})}\,, 
\label{eq:avers}
\end{equation}
which is derived again in sect.~4. The agreement is very good. Moreover, we see 
that $\E_{\scriptscriptstyle \text{FP}}[\phi^{(i)}_k]$ vanishes in
the thermodynamic limit for $i\ne k$, while it keeps finite for
$i=k$. The reason is that in this limit the vector $\phi$ includes
just one diagonal component per clique and an infinite number of
off--diagonal ones. 

\subsection{Distribution of the excess of votes}

\begin{figure}[!t]
  \begin{minipage}[!t]{0.5\textwidth}
  \centering
  \includegraphics[width=0.9\textwidth]{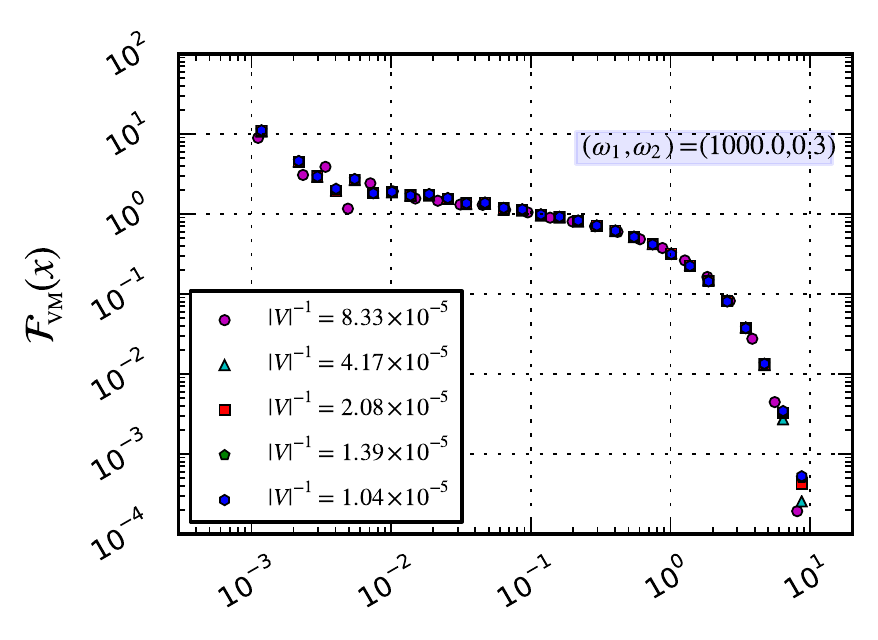}
  \vskip -0.3cm
  \includegraphics[width=0.9\textwidth]{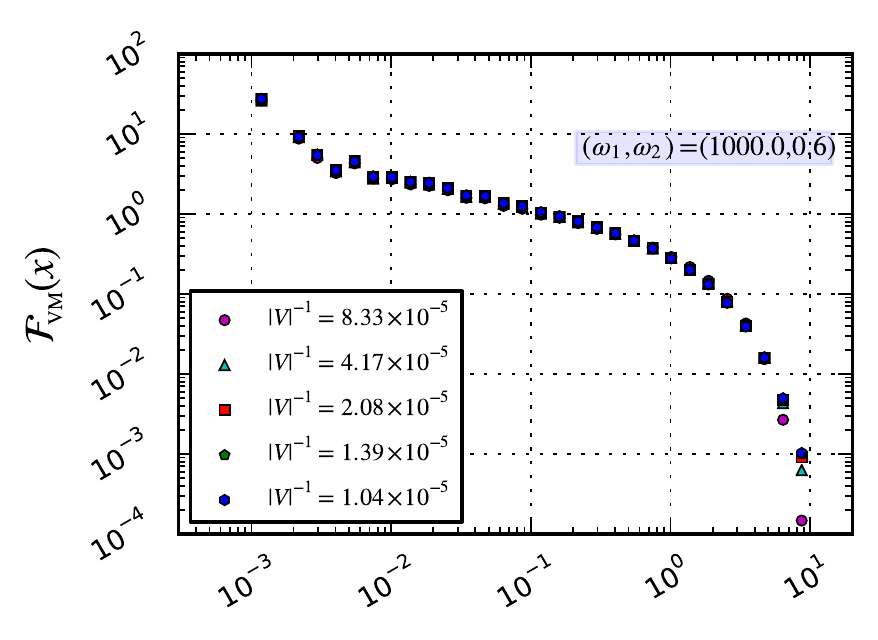}
  \vskip -0.3cm
  \includegraphics[width=0.9\textwidth]{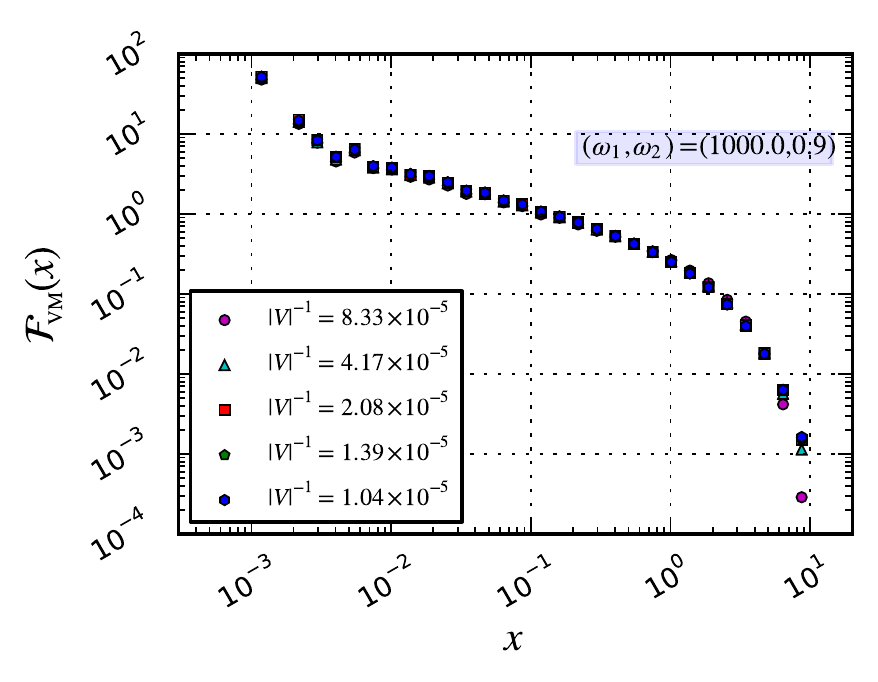}
  \end{minipage}
  \begin{minipage}[!t]{0.5\textwidth}
  \centering
  \includegraphics[width=0.9\textwidth]{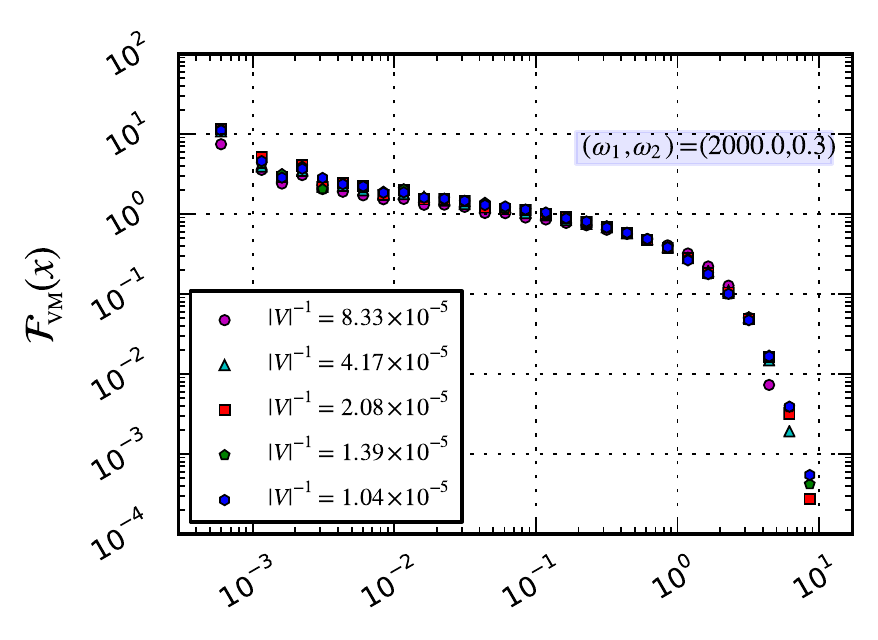}
  \vskip -0.3cm
  \includegraphics[width=0.9\textwidth]{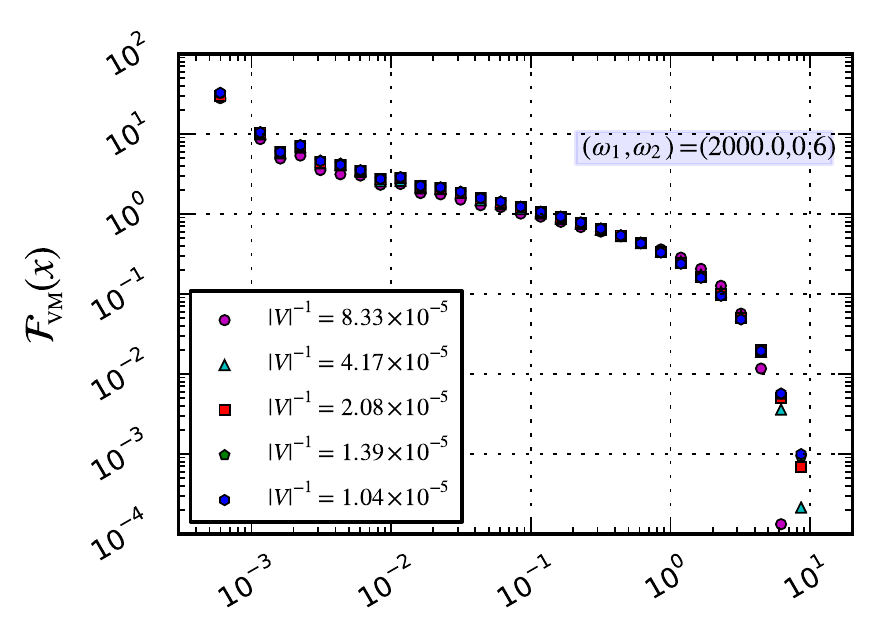}
  \vskip -0.3cm
  \includegraphics[width=0.9\textwidth]{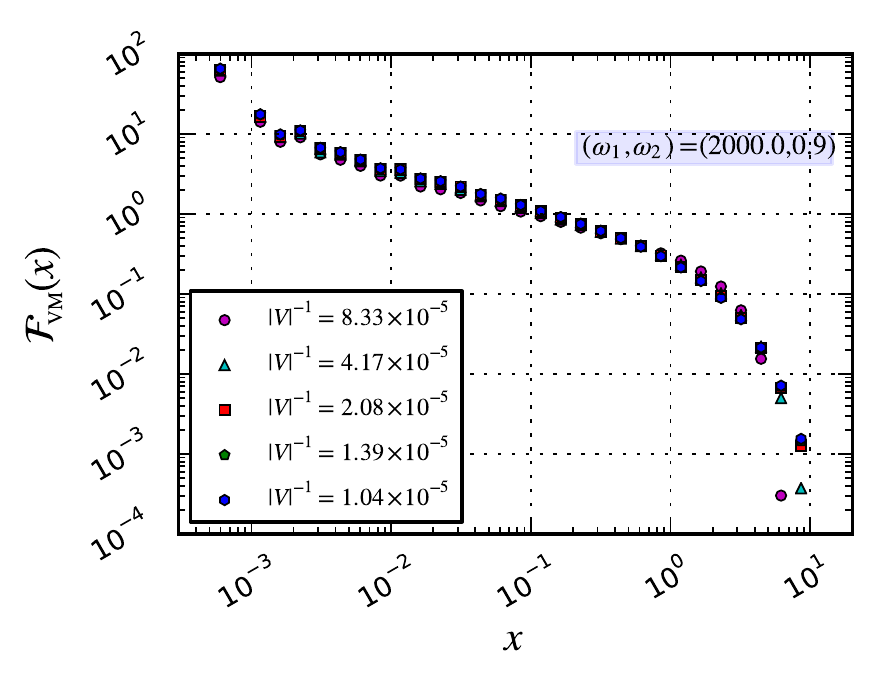}
  \end{minipage}
  \caption{\footnotesize Distribution of the intra--party excess of votes.}
  \label{fig:xs}
\end{figure}

In Fig.~\ref{fig:xs}, we show the distribution of the intra--party
excess of votes at all parameters of Table~\ref{tab:pars}. As
mentioned in sect.~1, plots resemble roughly the
distribution observed in the real Brazilian elections. A few considerations
are due. 
\begin{itemize}
\item{We notice at first glance that the dependence of the
    distributions upon $|V|$ is minimal: $\cFVM(x)$ is very close to the
    thermodynamic limit at all simulation parameters. Finite size
    effects are only visible along the right tails. This is in clear
    contrast with the substantial dependence upon $|V|$ displayed by
    the marginal distributions. Although the model bears the FC
    scaling formally just in the thermodynamic limit, in practice
    $\cFVM(x)$ scales very well also at finite $|V|$.} 
\item{The reader should be aware that the final lift of the left
    tails is fake. We have already observed in sect.~2.4 that
    $\phi_k$ is a discrete variable. Its quantization cannot be
    neglected at scales comparable to $\omega_1^{-1}$. In this
    region, a representation of $\cFVM(x)$ as a histogram is
    meaningless, as the number of events falling within any bin
    (surrounding a Dirac delta) is independent of the bin size,
    while the height of the corresponding rectangle is
    proportionally inverse to it. The histogram becomes stable at
    $\phi_k\gtrsim 5\omega_1^{-1}$, thus the last three to four
    points on the left should be ignored. We show them in
    Fig.~\ref{fig:xs} purposely to highlight the phenomenon.} 
\item{The central part of the distribution follows a power law
    $\cFVM(x)\propto x^{-\alpha}$, with $\alpha$ depending on the
    model parameters. In particular, $\alpha$ increases with
    $\omega_2$.}
\item{Unfortunately, the right tail of $\cFVM(x)$ is not
    log--normal, at least at the simulation parameters of Table~\ref{tab:pars}.  Since the
    distribution of the Brazilian data fits very well the FC
    universal curve  at $x\gtrsim 1$ (as pointed out in
    \cite{fempanal}), this appears to be a clear point of phenomenological
    weakness of our model.}  
\item{Establishing how $\cFVM(x)$ depends upon $(\omega_1,\omega_2)$
  is non--trivial. It would be tempting to introduce 
  renormalization group concepts by investigating whether $\cFVM(x)$ is
  invariant along some trajectory $\left(\omega_1,\omega_2(\omega_1)\right)$,
  but simulations are CPU demanding and we have not explored this
  possibility yet.}
\end{itemize}

In conclusion, upon evaluating the phenomenological valence of the model it
should be certainly kept in mind that $\cFVM(x)$ depends on just
two parameters, so the model should be just taken as a rough
prototype of more sophisticated (and realistic) constructions.
Moreover, $\cFVM(x)$ is in a sense a pure state distribution, in that
it does not mix different values of $(\omega_1,\omega_2)$. 

\section{Clique MFT}

Besides numerical simulations, MFT represents the simplest approach to 
studying birth--death models such as the VM. Its main benefit is the 
possibility of investigating the system in analytic 
terms. Although the Fokker--Planck equation of the model cannot be 
solved exactly (as we shall see below), it holds approximate 
information on the fluctuations of $\bar\phi$, 
on the dependence of  $\cPVM$ upon the network topology, on the influence 
exerted by each candidate on her own clique and the others and on how the competition 
between candidates leads to the specific equilibrium discussed in the previous 
section. Therefore, looking at the MFT of the model represents a 
necessary step towards a full understanding of the distribution of the
excess of votes.

In the framework of MFT, the system is described  in 
terms of macroscopic states and its probability density follows from a
Master Equation. The defining feature of MFT is the
observation of the microscopic dynamics from a distance comparable to the
size of the system. At this length scale, transition rates are
obtained by counting over the whole network the agents, who are
potentially responsible for the variation of macroscopic variables in a
given state. This approach is pursued 
in \cite{Mobilia2007} for the binary VM with zealots and in \cite{Starnini}
for the multi--state VM with no zealots. Its main drawback  is that
overall transition rates bring no information about the topological
details of the network. Thus, MFT fails to describe the system
whenever topology plays an essential r\^ole in the microscopic
dynamics. 

In the context of the model introduced in sect.~2, cliques are the leading
topological structures. Accordingly, it makes sense to increase the resolving
power of MFT by observing the microscopic dynamics from a distance comparable
to the size of a clique. To this aim, we observe that a single event
where the vote of $x\in V_0$ switches from $m$ to $\ell$, occurs nowhere but
the clique $C_i\ni x$. The transition turns the vote vector $v$
into a new one $v^{(i)}_{\ell^+m^-}$ with components
\begin{equation}
\left(v^{(i)}_{\ell^+m^-}\right)^{(r)}_s = \left\{\begin{array}{ll} v^{(i)}_s+1\,, & \text{if}
\ \ r=i\,,\quad s=\ell\,;\\[0.0ex]
v^{(i)}_s-1\,, & \text{if}
\ \ r=i\,,\quad s=m\,;\\[0.0ex]
v^{(i)}_s\,, & \text{if}
\ \ r=i\,,\quad s\ne \ell,m\,;\\[0.0ex]
v^{(r)}_s\,, & \text{if}
\ \ r\ne i\,, \quad s=1,\ldots,Q\,.
 \end{array}\right.
\end{equation}
At this length scale the Master Equation reads
\begin{equation}
\partial_t \cP(v)=\frac{1}{2}\sum_{i=1}^Q\sum_{\ell\ne m}^{1\ldots Q}
\left[w\left(v\bigr|v^{(i)}_{\ell^+ m^-}\right)\cP(v^{(i)}_{\ell^+ m^-}) -
  w\left(v^{(i)}_{\ell^+ m^-}\bigr|v\right)\cP(v) \right]\,,
\label{eq:master}
\end{equation}
where $\partial_t=\partial/\partial t$ and $w(v^{(i)}_{\ell^+ m^-}|v)$ measures the 
rate of the transition $v\to v^{(i)}_{\ell^+ m^-}$. We then notice that
$w$ is made of two contributions, describing events due to
intra-- and  inter--clique interactions. Specifically, we have
\begin{equation}
w\left(v^{(i)}_{\ell^+ m^-}\bigr|v\right) = \frac{1}{\Delta
  t}\frac{v^{(i)}_m-\delta_{mi}}{\dfrac{|V|}{Q}-1}\left[\pi_\text{in}
  \frac{v^{(i)}_\ell}{\dfrac{|V|}{Q}-1}  
  + 
  \frac{\pi_\text{out}}{Q-1}\sum_{k\ne i}^{1\ldots
    Q}\frac{v^{(k)}_\ell-\delta_{kl}}{\dfrac{|V|}{Q}-1}\right]\,, 
\label{eq:transrate}
\end{equation}
\begin{equation}
w\left(v\bigr|v^{(i)}_{\ell^+ m^-}\right) = \frac{1}{\Delta
  t}\frac{v^{(i)}_\ell+1-\delta_{\ell
    i}}{\dfrac{|V|}{Q}-1}\left[\pi_\text{in}\frac{v^{(i)}_m-1}{\dfrac{|V|}{Q}-1}
  + \frac{\pi_\text{out}}{Q-1}\sum_{k\ne i}^{1\ldots
    Q}\frac{v^{(k)}_m-1-\delta_{km}}{\dfrac{|V|}{Q}-1}\right]\,. 
\label{eq:invtransrate}
\end{equation}
To check the above expressions, the reader should keep in mind that: $i$) the
only agents experiencing transitions are those in $V_0$; $ii$)~candidates
participate in the microscopic dynamics as opinion donors; $iii$)~candidates
do not take part in inter--clique interactions.  

It should be also observed that, since we measure time in units of sweeps, an
elementary transition occurs on average in a time interval $\delta
t = (|V|-Q)^{-1}$. Since the network is made of $Q$ cliques, a
transition within a specific one occurs on average in a time
interval which is larger than $\delta t$ by a factor of
$Q$. Accordingly, we set $\Delta t = Q\delta t =  Q/(|V|-Q) =
1/\delta_{\text{in}}$.  

Thanks to the definitions introduced in eq.~(\ref{eq:deltainout}),
eqs.~(\ref{eq:transrate})--(\ref{eq:invtransrate}) simplify to
\begin{equation}
w\left(v^{(i)}_{\ell^+ m^-}|v\right) = \left(v^{(i)}_m
  -\delta_{mi}\right)\left[\frac{\pi_\text{in}}{\delta_\text{in}}v^{(i)}_\ell
  + p\frac{\pi_\text{out}}{\delta_\text{out}}\sum_{k\ne
    i}^{1\ldots Q}(v^{(k)}_\ell-\delta_{k\ell})\right]\,, 
\end{equation}
\begin{equation}
w\left(v\bigr|v^{(i)}_{\ell^+ m^-}\right) = \left(v^{(i)}_\ell +
  1-\delta_{\ell
    i}\right)\left[\frac{\pi_\text{in}}{\delta_\text{in}}(v^{(i)}_m-1)
  + p\frac{\pi_\text{out}}{\delta_\text{out}}\sum_{k\ne
    i}^{1\ldots Q}(v^{(k)}_m-1-\delta_{km})\right]\,. 
\end{equation}

\subsection{Kramers--Moyal expansion}

Seldom can the Master Equation be solved analytically. Indeed,
eq.~(\ref{eq:master}) is of no exception. An approximation to it
can be obtained by means of the Kramers--Moyal expansion, which
upon truncation to second order yields the FP equation (see for
instance ch. 7 of \cite{Gardiner}). To work out the approximation,
we preliminarily express the transition rates in terms of the
scaling variables and then expand eq.~(\ref{eq:master}) in Taylor
series at the target point $\bar\phi$, where we wish to calculate
the probability density. It is worth noticing some subtleties: 
\begin{itemize}
\item{The expansion parameter is identified here with the minimum variation of
  $\phi^{(i)}_k$, namely $\Delta\phi^{(i)}_k=\omega_1^{-1}$, while in 
  \cite{Starnini,Mobilia2007} it is given by $|V|^{-1}$. In the
  thermodynamic limit the former stays finite whereas the latter
  vanishes. Therefore, the Kramers--Moyal expansion we perform in this
  paper is somewhat different than in standard MFT. Anyway, the Taylor
  series is expected to converge rapidly, provided $\omega_1 \gg 1$.}
\item{Differentiating the probability density with respect to $\phi^{(i)}_k$
  is only possible provided the latter is assumed to be a continuous 
  variable. Such assumption is certainly correct at
  $\phi^{(i)}_k\gg\omega_1^{-1}$, while it fails miserably close
  to the quantization scale. For this reason, we expect that
  distributions obtained from the FP equation deviate from the VM 
  at scales comparable to $\omega_1^{-1}$.} 
\item{In the thermodynamic limit, it holds $\Delta t =
  (\omega_1-1)^{-1}$. Thus, we see that time is also
  quantized. Nevertheless, we treat time as a continuous variable
  under the assumption $\omega_1\gg 1$.}   
\end{itemize}
Having said that, a simple rescaling of the transition rates yields
\begin{align}
& \tilde w\left(\phi^{(i)}_{\ell^+ m^-}\bigr|\phi\right) =
\omega_1\frac{\phi^{(i)}_m
  -\omega_1^{-1}\delta_{mi}}{(1-\omega_1^{-1})(1+\omega_2)}\left[\phi^{(i)}_\ell
  + \frac{\omega_2}{Q-1}\sum_{k\ne i}^{1\ldots
    Q}\left(\phi^{(k)}_\ell -
    \omega_1^{-1}\delta_{k\ell}\right)\right]\,,\\[0.0ex] & \tilde
w\left(\phi\bigr|\phi^{(i)}_{\ell^+ m^-}\right) =
\omega_1\frac{\phi^{(i)}_\ell +\omega_1^{-1}(1-\delta_{\ell
    i})}{(1-\omega_1^{-1})(1+\omega_2)}\left[\phi^{(i)}_m-\omega_1^{-1}\right. 
  + \left.\frac{\omega_2}{Q-1}\sum_{k\ne i}^{1\ldots Q}\left(\phi^{(k)}_m -
  \omega_1^{-1}(1+\delta_{km})\right)\right]\,,
\end{align}
with $\tilde w\left(\phi^{(i)}_{\ell^+
  m^-}\bigr|\phi\right) \equiv w\left(v^{(i)}_{\ell^+
  m^-}\bigr|v\right)$. The result of the Kramers--Moyal expansion
is the FP equation 
\begin{equation}
\partial_t\cPFP(\bar\phi) = -\sum_{i=1}^Q\sum_{\ell\ne i}^{1\ldots
  Q}\partial^{(i)}_\ell
\left[A^{(i)}_\ell(\bar\phi)\cPFP(\bar\phi)\right] +
\frac{1}{2}\sum_{i,j=1}^{Q}\sum_{\ell,m\ne i}^{1\ldots
  Q}\partial^{(i)}_\ell\partial^{(j)}_{\ml}\left[B^{(ij)}_{\ell
    m}(\bar\phi)\cPFP(\bar\phi)\right]\,, 
\label{eq:fpeq}
\end{equation}
where $\partial^{(i)}_\ell = \partial/\partial\phi^{(i)}_\ell$. Not
surprisingly, the analytic structure of eq.~(\ref{eq:fpeq}) reflects the
specific topology of the network. The coefficient functions $A^{(i)}_\ell$ and
$B^{(ij)}_{\ell m}$ depend on the transition rates via the standard formulae
\begin{align}
A^{(i)}_\ell & = \frac{1}{2}\sum_{k=1}^Q\sum_{a\ne b}^{1\ldots
  Q}r\left(\phi\to\phi^{(k)}_{a^+b^-}\right)^{(i)}_\ell
\alpha^{(k)}_{a^+b^-}\,,\\[0.0ex] B^{(ij)}_{\ell m} & =
\frac{1}{2}\sum_{k=1}^Q\sum_{a\ne b}^{1\ldots
  Q}r\left(\phi\to\phi^{(k)}_{a^+b^-}\right)^{(i)}_\ell
r\left(\phi\to\phi^{(k)}_{a^+b^-}\right)^{(j)}_m
\beta^{(k)}_{a^+b^-}\,, 
\end{align}
with
\begin{align}
\alpha^{(k)}_{a^+b^-} & = \tilde w\left(\phi^{(i)}_{a^+
    b^-}\bigr|\phi\right) - \tilde w\left(\phi^{(i)}_{a^-
    b^+}\bigr|\phi\right)\,,\\[0.0ex] \beta^{(k)}_{a^+b^-} & =
\tilde w\left(\phi^{(i)}_{a^+ b^-}\bigr|\phi\right) + \tilde
w\left(\phi^{(i)}_{a^- b^+}\bigr|\phi\right)\,, 
\end{align}
and
\begin{equation}
r\left(\phi\to\phi^{(k)}_{a^+b^-}\right)^{(i)}_\ell =
\frac{\delta_{ki}}{\omega_1}\left(\delta_{\ell a} - \delta_{\ell
    b}\right)\,. 
\end{equation}
Owing to the symmetries $\alpha^{(k)}_{a^+b^-} = - \alpha^{(k)}_{b^+a^-}$ and
$\beta^{(k)}_{a^+b^-} = \beta^{(k)}_{b^+a^-}$, the above expressions can be
brought to the simplified form 
\begin{align}
A^{(i)}_\ell & = \frac{1}{\omega_1}\sum_{a\ne \ell}^{1\ldots
  Q}\alpha^{(i)}_{\ell^+a^-}\,,\\[0.0ex] 
\label{eq:Bsimp} B^{(ij)}_{\ell m} & =
\frac{\delta_{ij}}{\omega_1^2}\left[\delta_{\ell m}\sum_{a\ne
    \ell}^{1\ldots Q}\beta^{(i)}_{\ell^+a^-} - (1-\delta_{\ell
    m})\beta^{(i)}_{\ell^+m^-}\right] \ \equiv\
\delta_{ij}B^{(i)}_{\ell m} \,. 
\end{align}
We thus see that $B^{(ij)}_{\ell m}$ is diagonal with respect to the clique
indices $(ij)$. Yet, we shall show in a moment that $B^{(i)}_{\ell
  m}$ depends upon all the components of $\bar\phi$ and not just on
$\{\phi^{(i)}_k\}_{k\ne i}$. In addition we see that
$B^{(i)}_{\ell m}$ is symmetric with respect to the indices $(\ell 
m)$. Consequently, $\cPFP(\bar\phi)$ is invariant under the transformations
\begin{alignat}{4}
\label{eq:intrasym}& \text{intra--clique symmetry:}\qquad &
\phi^{(i)}_\ell &\, \longleftrightarrow\, \phi^{(i)}_m\,,\qquad &
& i=1,\ldots, Q,\quad \ell,m\ne i\,,\\[2.0ex] 
\label{eq:intersym}& \text{inter--clique symmetry:}\qquad &
\{\phi^{(i)}_k\}_{k\ne i}^{1\ldots Q} &\, \longleftrightarrow\,
\{\phi^{(j)}_k\}_{k\ne j}^{1\ldots Q}\,,\qquad & & i,j
=1,\ldots,Q\,.
\end{alignat}
We finally observe that at equilibrium eq.~(\ref{eq:fpeq}) amounts to the
local conservation law
\begin{equation}
0=\sum_{i=1}^Q\sum_{\ell\ne i}^{1\ldots
  Q}\partial^{(i)}_{\ell}J^{(i)}_\ell(\bar\phi)\,, 
\label{eq:conslaw}
\end{equation}
with the vector current $J^{(i)}_\ell(\bar\phi)$ given by
\begin{equation}
J^{(i)}_\ell(\bar\phi) =
A^{(i)}_\ell(\bar\phi)\cPFP(\bar\phi)-\frac{1}{2}\sum_{m\ne
  i}^{1\ldots Q}\partial^{(i)}_{\ml}\left[B^{(i)}_{\ell
    m}(\bar\phi)\cPFP(\bar\phi)\right]\,. 
\label{eq:Jfield}
\end{equation}
In the context of the binary VM with zealots \cite{Mobilia2007}, an
exact solution of the FP equation has been obtained by just using
the conserved current and a proportionality relation between the
coefficient functions. In the present case, proceeding along the
same lines looks rather hard. Indeed, since the problem has more
than one independent variable, eq.~(\ref{eq:conslaw}) does not
entail $J^{(i)}_\ell=\text{const}$. In addition, the relation
between $A^{(i)}_\ell$ and $B^{(i)}_{\ell m}$ is more
complicated. Yet, we cannot exclude that smart tricks, which
unfortunately we have not been able to find, do exist to solve the
FP equation analytically.  

It takes one page of scratch paper to work out the coefficient
functions. A convenient representation, easy to implement on
a computer, is given by 
\begin{align}
\tauFP A^{(i)}_\ell & =
-\left[1+\omega_1\omega_2(1-\omega_1^{-1})\right]\phi^{(i)}_\ell +
\frac{\omega_1\omega_2(1-\omega_1^{-1})}{Q-1}\left\{\sum_{k\ne
    i,\ell}^{1\ldots Q}\phi^{(k)}_\ell - \sum_{k\ne\ell}^{1\ldots
    Q}\phi^{(\ell)}_k\right\}\nonumber\\[2.0ex]
&  + \frac{\omega_1\omega_2(1-\omega_1^{-1})^2}{Q-1}\,,
\label{eq:driftone}
\end{align}
\begin{align}
\tauFP B^{(i)}_{\ell m} = -2(1-\delta_{\ell
  m})\phi^{(i)}_\ell\phi^{(i)}_m - (1-\delta_{\ell
  m})\frac{\omega_2}{(Q-1)}&
\left[\phi^{(i)}_\ell\left(1-\omega_1^{-1} + \sum_{k\ne
      i,m}^{1\ldots Q} \phi^{(k)}_m - \sum_{k\ne m}^{1\ldots Q\ 
}\phi^{(m)}_k\right)\right.\nonumber\\[1.0ex] 
& \hskip -1.5ex + \left.\phi^{(i)}_m\left(1-\omega_1^{-1} +
    \sum_{k\ne i,\ell}^{1\ldots Q} \phi^{(k)}_\ell - \sum_{k\ne
      \ell}^{1\ldots Q}\phi^{(\ell)}_k\right)\right]\nonumber 
\end{align}
\vskip -0.5cm
\begin{align}
\phantom{\tau B^{(i)}_{\ell m}}& + 2\delta_{\ell
  m}\phi^{(i)}_\ell\left[1 +
  \frac{\omega_2(1-\omega_1^{-1})-\omega_1^{-1}}{2}-\phi^{(i)}_\ell\right]
\nonumber\\[1.0ex]  
& + \delta_{\ell
  m}\frac{\omega_2}{(Q-1)}\left(1-\omega_1^{-1}-2\phi^{(i)}_\ell\right)\left[1
  - \omega_1^{-1} + \sum_{k\ne i,\ell}^{1\ldots Q}\phi^{(k)}_\ell
  - \sum_{k\ne\ell}\phi^{(\ell)}_k\right]\,, \qquad \ell,m\ne
i\,,
\label{eq:diffusone}
\end{align}
with $\tauFP$ having been defined in eq.~(\ref{eq:tauFP}). We see
that as $\omega_1\to\infty$ and $\omega_2\to 0$, the 
diffusive coefficient function collapses to the expression found
in~\cite{Starnini}. The fate of the drift term in the
same limit depends on the behavior of the product
$\omega_1\omega_2$.

\subsection{Autocorrelation time and average values from clique MFT}

The drift equation obtained by deleting the diffusion term in
eq.~(\ref{eq:fpeq}) is equivalent to a system of coupled linear differential
equations, 
\begin{equation}
\frac{\ \rd\varphi^{(i)}_\ell}{\rd t} =
A^{(i)}_\ell(\bar\varphi)\,,
\label{eq:FPaver}
\end{equation}
with $\varphi^{(i)}_\ell = \E_{\scriptscriptstyle
  \text{FP}}[\phi^{(i)}_\ell]$. The expression given in
eq.~(\ref{eq:driftone}) for the drift coefficient  has been
obtained upon applying the simplex constraints, thus it assumes
$\ell\ne i$. A more general expression, less compact but valid
with no  restrictions, is 
\begin{align}
\tauFP A^{(i)}_\ell(\bar\varphi) & = \sum_{m\ne\ell}^{1\ldots Q}\left(\delta_{\ell
    i}\varphi^{(i)}_m - \delta_{mi}\varphi^{(i)}_\ell\right)
\nonumber\\[1.0ex] 
& \hskip -0.8cm +
\frac{\omega_1\omega_2}{Q-1}\sum_{m\ne\ell}^{1\ldots
  Q}\sum_{k\ne i}^{1\ldots
  Q}\left[(\varphi^{(i)}_m-\omega_1^{-1}\delta_{mi})(\varphi^{(k)}_\ell
  - \omega_1^{-1}\delta_{k\ell}) -
  (\varphi^{(i)}_\ell-\omega_1^{-1}\delta_{\ell i})(\varphi^{(k)}_m -
  \omega_1^{-1}\delta_{km}) \right]\,. 
\label{eq:meanevol}
\end{align}
We note that the term in square brackets on the second line is antisymmetric
in the clique indices~$i,k$. Upon summing both sides of
eq.~(\ref{eq:FPaver}) over $i$, that term averages to zero. By
using the identities $\sum_{m\ne \ell}\delta_{m i}=(1-\delta_{\ell
  i})$ and $\sum_{m\ne\ell}\varphi^{(i)}_m = 1-\varphi^{(i)}_\ell$, we
obtain 
\begin{equation}
\tauFP \frac{\rd\varphi_\ell}{\rd t} = (1-\varphi_\ell)\,.
\end{equation}
This equation is trivially solved by the function
\begin{equation}
\varphi_\ell(t) = 1 + (\varphi_\ell(0) - 1)\re^{-t/\tauFP}\,.
\label{eq:equiphi}
\end{equation}
Since the analytic form of $\varphi_\ell(t)$ is now known, we can
use $\sum_{k\ne i}\varphi^{(k)}_\ell(t) = \varphi_\ell(t) -
\varphi^{(i)}_\ell(t)$ to reduce all terms of eq.~(\ref{eq:FPaver})
to functions of the unknown $\varphi^{(i)}_\ell(t)$ alone. A little
algebra  
yields
\begin{equation}
\tauFP\frac{\rd\varphi_\ell^{(i)}}{\rd t} = \alpha\varphi_\ell -
\beta\varphi^{(i)}_\ell + \gamma_{i\ell}\,, 
\label{eq:simpevol}
\end{equation}
with
\begin{align}
\alpha & = \frac{\omega_2(\omega_1-1)}{Q-1}\,,\\[0.0ex] 
\beta  & = 1+\omega_2(\omega_1-1)\frac{Q}{Q-1}\,,\\[0.0ex]
\gamma_{i\ell} & =
\delta_{i\ell}\left[1+\omega_2(1-\omega_1^{-1})\frac{Q}{Q-1}
\right]-\frac{\omega_2(1-\omega_1^{-1})}{Q-1}\,.  
\end{align}
Eq.~(\ref{eq:simpevol}) can be integrated exactly. The trick to do that is to
multiply both sides of it by the integrating factor $\exp(\beta t/\tauFP)$ and to
observe that
\begin{equation}
\tauFP\re^{\beta t/\tauFP}\frac{\rd\varphi^{(i)}_\ell}{\rd t} = \tauFP\frac{\rd}{\rd
  t}\left(\re^{\beta t/\tauFP}\varphi^{(i)}_\ell\right) -
\beta\re^{\beta t/\tauFP}\varphi^{(i)}_\ell\,. 
\end{equation}
Since the indefinite integral of $\varphi_\ell(t)$  is known, the solution to
eq.~(\ref{eq:simpevol}) is given by
\begin{equation}
\varphi^{(i)}_\ell(t) = \re^{-\beta t/\tauFP}\varphi^{(i)}_\ell(0) +
\frac{\alpha+\gamma_{i\ell}}{\beta}\left(1-\re^{-\beta t/\tauFP}\right) + \frac{
  \alpha\left[\varphi_\ell(0)-1\right]}{\beta - 1}\left(\re^{-t/\tauFP} -
\re^{-\beta t/\tauFP}\right)\,.
\end{equation}
We observe that $\beta\gg 1$ at all simulation parameters of
Table~\ref{tab:pars}. Therefore, the time evolution of the
average variables is dominated by two largely separated time scales. $\tauFP$
is the slowest mode: it governs the relaxation to equilibrium of
the averages or --- equivalently --- the autocorrelation
time of the model. In particular, we have 
\begin{equation}
\lim_{t\to\infty}\varphi^{(i)}_\ell(t) = \frac{\alpha+\gamma_{i\ell}}{\beta} =
\frac{\delta_{i\ell}\left[Q-1+\omega_2
    Q(1-\omega_1^{-1})\right]+\omega_1\omega_2(1-\omega_1^{-1})^2}{Q-1
  +\omega_1\omega_2Q(1-\omega_1^{-1})}\,, 
\label{eq:phieq}
\end{equation}
and
\begin{align}
& \lim_{|V|\to\infty}\lim_{t\to\infty}\varphi^{(i)}_i(t) =
\frac{1+\omega_2(1-\omega_1^{-1})}{1+\omega_1\omega_2(1-\omega_1^{-1})}\equiv
\varphi_\diag(\omega_1,\omega_2)\,,\\[2.0ex] 
&
\lim_{|V|\to\infty}\lim_{t\to\infty}(Q-1)\varphi^{(i)}_\ell(t)\bigr|_{i\ne\ell}
=
\frac{\omega_1\omega_2(1-\omega_1^{-1})^2}{1+\omega_1\omega_2(1-\omega_1^{-1})}
\equiv\varphi_\off(\omega_1,\omega_2)\,. 
\end{align}
Obviously, $\varphi_\diag+\varphi_\off=1$. 

\section{Numerical integration of the FP equation}

It is well known that a multidimensional FP equation with
probability density confined within a convex domain $\cD$ is equivalent to 
a set of coupled stochastic differential equations with reflecting
boundary conditions (b.c.) on $\partial\cD$. In the specific case of
eq.~(\ref{eq:fpeq}) with b.c. $\cPFP(\bar\phi)=0$ if
$\phi\notin\cT_Q(\omega_1^{-1})$, the equivalent stochastic process is
given by 
\begin{equation}
\rd\phi^{(i)}_\ell(t) = A^{(i)}_\ell(\bar\phi)\,\rd t + \sum_{k\ne
  i}^{1\ldots Q}C^{(i)}_{\ell k}(\bar\phi)\,\rd W^{(i)}_k(t) +
\rd K^{(i)}_\ell(t)\,,\qquad i=1,\ldots,Q\,,\quad \ell\ne i\,.
\label{eq:ito}
\end{equation}
with the matrix $C^{(i)}_{\ell k}$ being related to $B^{(i)}_{\ell k}$ via
\begin{equation}
B^{(i)}(\bar\phi) =
C^{(i)}(\bar\phi)\cdot\trans{[C^{(i)}(\bar\phi)]}\,.
\end{equation}
As usual, $W(t) = \{W^{(i)}_k(t)\}_{i\ne k}^{1\ldots Q}$ is a
multidimensional Wiener process, while
$K(t)=\{K^{(i)}_k(t)\}_{i\ne k}^{1\ldots Q}$ is a 
multidimensional bounded variation process (known as
a Skorokhod term), increasing only when  \linebreak
$\phi\in\partial\cT_Q(\omega_1^{-1})$.  
\begin{figure}[t!]
    \centering
    \includegraphics[width=0.9\textwidth]{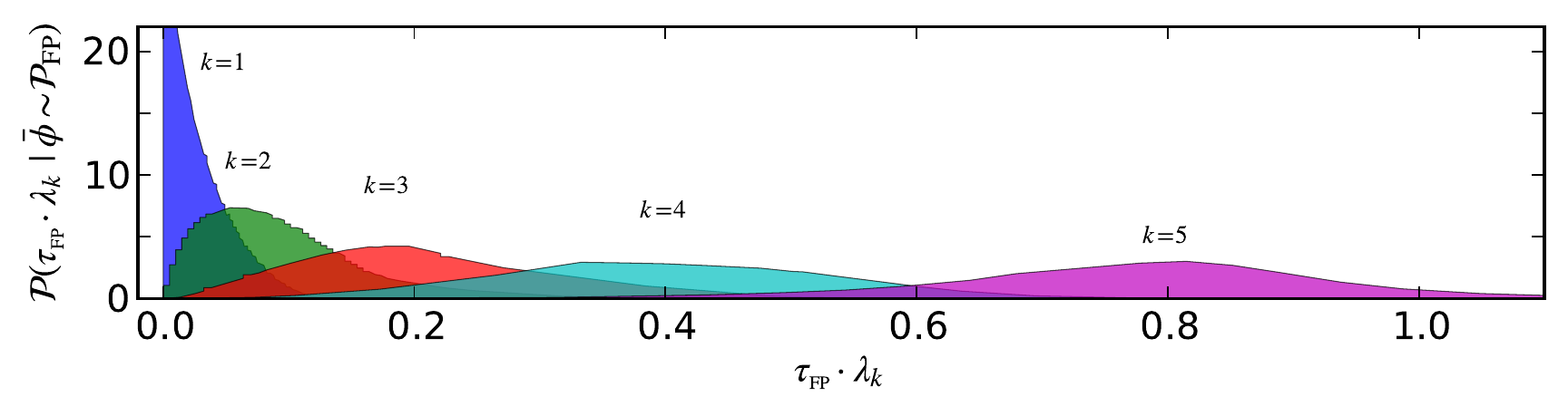}
  \caption{ \footnotesize Probability density of the ordered eigenvalue
    spectrum of $B^{(i)}$ conditioned to $\bar\phi\sim\cPFP$ at
    $(\omega_1,\omega_2)=(2,\!000.0,0.3)$ and $Q=6$, as obtained 
  from numerically integrating eq.~(\ref{eq:ito}).}
  \label{fig:eva} 
\end{figure}
In this section, we report on some experiences of numerical integration
of eq.~(\ref{eq:ito}), performed via the classical Euler
integration scheme, first considered in \cite{Maruyama}. Technical
details follow. 

First of all, we obtain the matrix $C^{(i)}$ numerically by
performing a Cholesky factorization of $B^{(i)}$ at each time
integration step. This requires $B^{(i)}(\bar\phi)$ to be positive
definite for $\phi\in\cT_Q(\omega_1^{-1})$. Since we lack an
analytic proof of this property, we check it numerically. Owing
to the inter--clique symmetry given by eq.~(\ref{eq:intersym}), the
distribution of the eigenvalue spectrum $\lambda = \{\lambda_k\}_{k=1}^{Q-1}$ of
$B^{(i)}$ is actually independent of $i$. We assume here $\lambda$ to be
increasingly ordered. In Fig.~\ref{fig:eva}, we show the
distribution of $\tauFP\cdot \lambda$ conditioned to $\bar\phi\sim \cPFP$ at 
$(\omega_1,\omega_2)=(2,\!000.0,0.3)$ and $Q=6$, as obtained by numerically
integrating eq.~(\ref{eq:ito}) with $\rd t=0.001$ and b.c. type 2
(see below).

Secondly,  in sect.~4 we proved that in absence of the
Wiener term, the autocorrelation time of eq.~(\ref{eq:ito}) 
equals $\tauFP$. We have checked numerically that upon switching 
the diffusion term on, the autocorrelation time does not change 
significantly. Therefore, along all numerical integrations we
retain only configurations of $\bar\phi$ separated by a number of
integration steps  
\begin{equation} 
\Delta N = k\frac{\tauFP}{\rd t}\,,
\label{eq:deltaN}
\end{equation}
with $k=3$ to $4$. Some experience shows that integrating the FP
equation numerically, with $\Delta N$ defined as in
eq.~(\ref{eq:deltaN}), is roughly as much CPU 
demanding as is simulating the VM. 

Finally, we observe that depending on the precise definition of
$K(t)$, the equilibrium distribution of $\bar\phi(t)$ at
finite $\rd t$ may be more or less close to
$\cPFP(\bar\phi)$. This problem has been investigated in general 
terms in~\cite{Slominski}, where a normal reflection scheme at the
boundary is adopted and the rate of $L^p$ convergence of the
stochastic process as $\rd t\to 0$ is discussed in full
detail. Unfortunately, normal reflections are ill--defined on 
$\partial\cT_Q(\omega_1^{-1})$ due to the edges of the simplices,
thus we cannot follow \cite{Slominski}. Instead, we consider two
algorithmic definitions of $K(t)$. By letting $\bar\phi\to\bar\phi'$
denote the result of a given Euler integration step with
$\phi\in\cT_Q(\omega_1^{-1})$, we define 

\begin{itemize}
\item{{\bf b.c. type 1} (accept--reject):}
  {
    \begin{itemize}
      {
        \setlength\itemindent{15pt}
      \item[\qquad {\it step 1})]{\quad if $(\phi')^{(i)}_\ell <0$ for some $i,\ell$,
          then $\bar\phi'$ is rejected;}
      \item[\qquad {\it step 2})]{\quad if $\sum_{\ell\ne i}^{1\ldots Q}(\phi')^{(i)}_\ell
          > 1-\omega_1^{-1}$ for some $i$, then $\bar\phi'$ is
          rejected.}
      }
    \end{itemize}
  }
  \vskip 0.2cm
\item{{\bf b.c. type 2} (boundary projection):}
  \begin{itemize}
    {
      \setlength\itemindent{15pt}
    \item[{\it step 1})]{\quad if $(\phi')^{(i)}_\ell <0$ for some $i,\ell$,
        then we set $(\phi')^{(i)}_\ell \gets 0$ ;}
    \item[{\it step 2})]{\quad if $\sum_{\ell\ne i}^{1\ldots Q}(\phi')^{(i)}_\ell
        > 1-\omega_1^{-1}$ for some $i$, then we set
        $(\phi')^{(i)}_\ell \gets c\cdot(\phi')^{(i)}_\ell$, with}
    }
  \end{itemize}
\end{itemize}
\begin{equation}
  c = \frac{1-\omega_1^{-1}}{\sum_{\ell\ne i}^{1\ldots Q}(\phi')^{(i)}_\ell}\,.
\end{equation}

\begin{figure}[t!]
  \begin{minipage}{0.5\textwidth}
  \centering
  \includegraphics[width=0.9\textwidth]{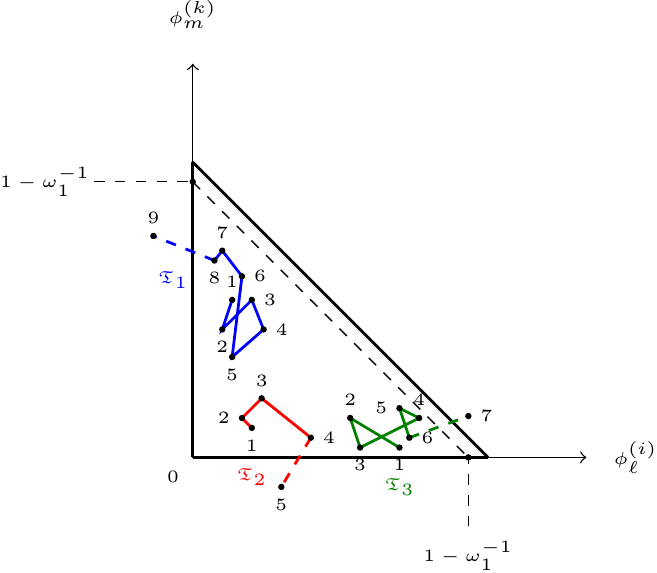}
  \end{minipage}
  \begin{minipage}{0.5\textwidth}
  \centering
  \includegraphics[width=0.9\textwidth]{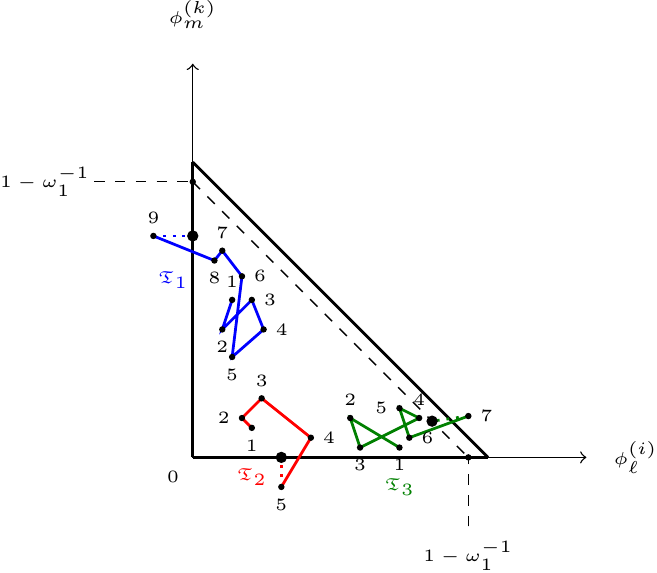}
  \end{minipage}
  \vskip -0.2cm
  \caption{\footnotesize {\it Left}: b.c. type 1. The plot shows
    schematically three trajectories ${\frak T}_k$ ($k=1,2,3$) crossing
  $\partial \cT_Q(\omega_1^{-1})$. The integration steps $8\to 9$ of ${\frak
    T}_1$, $4\to 5$ of ${\frak T}_2$ and $6\to 7$ of ${\frak T}_3$
  are all rejected. {\it Right}: b.c. type 2. The same trajectories
  ${\frak T}_k$ are processed differently. Points $9$ of ${\frak
    T}_1$, $5$ of ${\frak T}_2$ and $7$ of ${\frak T}_3$ are
  projected onto the boundaries. Both plots assume $m\ne k$ and
  $\ell\ne i$.}
  \label{fig:bc} 
  \vskip 0.3cm
\end{figure}

\noindent In Fig.~\ref{fig:bc} we show schematically some examples
of trajectories 
crossing $\partial\cT_Q(\omega_1^{-1})$ at finite $\rd t$. The
plots show how the crossings are handled according to the above
recipes. It is easily recognized that b.c. type 1 disfavor
updates $\bar\phi\to\bar\phi'$ with $\bar\phi'$ falling closer than
$\text{O}(\sqrt{\rd t})$ to the boundary (recall that $\rd t$ is the
variance of the increments of a Wiener process). By contrast,
b.c. type 2 enhance the r\^ole of the  boundary at finite $\rd
t$. We shall see the consequences of this in a moment.

\begin{figure}[t!]
  \vskip 2.0cm
    \centering
    \includegraphics[width=0.7\textwidth]{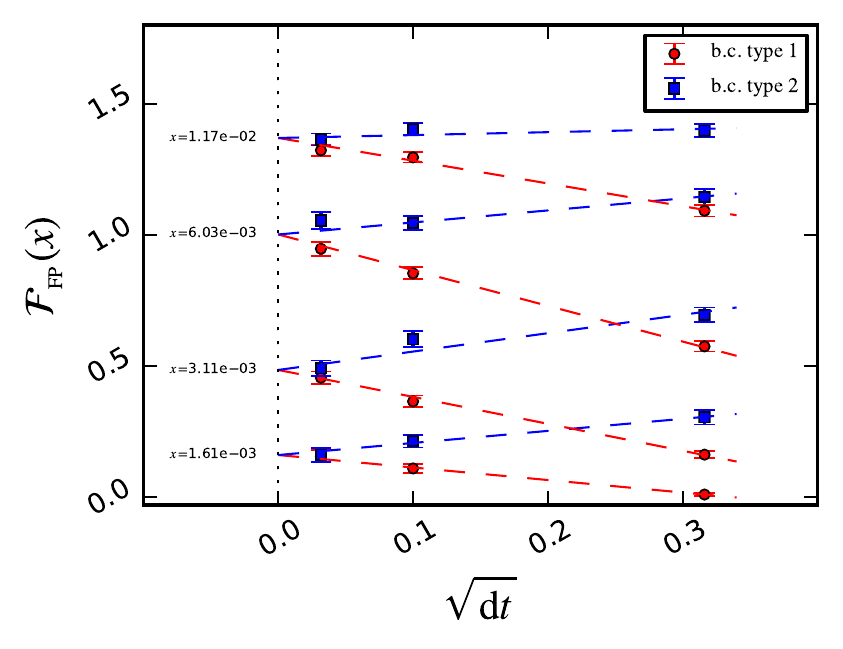}
    \vskip 1.0cm
    \includegraphics[width=0.7\textwidth]{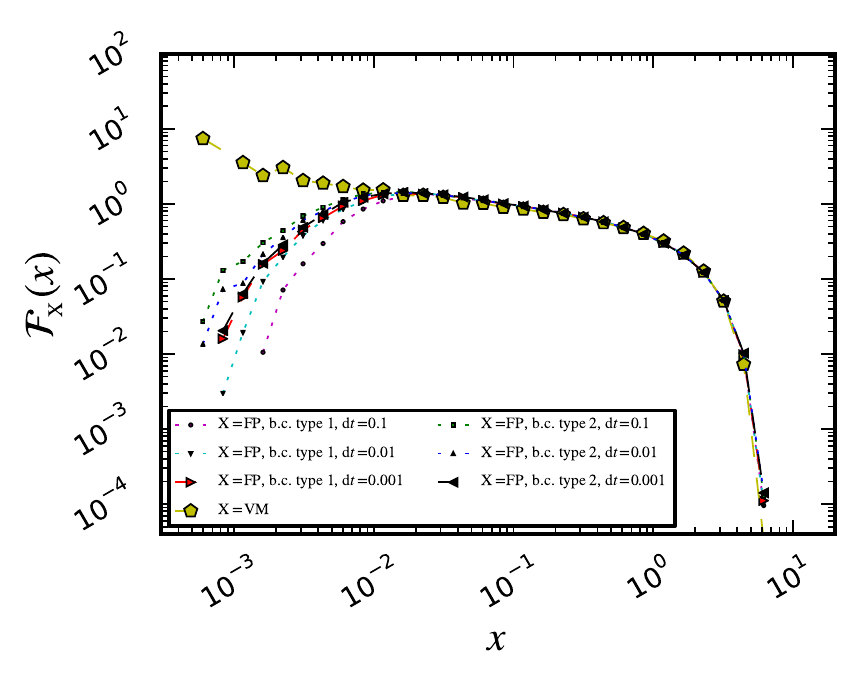}
  \caption{ \footnotesize {\it Top}: convergence of $\cFFP(x)$ with
    b.c. type 1 and type 2 as $\rd t\to 0$ for some values of
    $x$. Data refer to parameters
    $(\omega_1,\omega_2)=(2,\!000.0,0.3)$ and $Q=6$. 
  {\it Bottom}: Comparison between $\cFVM(x)$ and $\cFFP(x)$ as obtained
  by numerical integration of eq.~(\ref{eq:ito}) for several
  values of $\rd t$. The plot refers to the same choice
  of parameters as above.}
  \label{fig:fpdistr} 
\end{figure}

In Fig.~\ref{fig:fpdistr} (top), we show examples of
convergence of the distribution of the excess of votes $\cFFP(x)$
with b.c. type 1 and type 2 as $\rd t\to 0$ for some values of
$x$. Data points refer to numerical integrations performed with
$\rd t=0.1,\,0.01,\,0.001$ at $(\omega_1,\omega_2)=(2,\!000.0,0.3)$
and $Q=6$. Statistical uncertainties have been obtained by
bootstrapping new distributions from those simulated, with the bin
size adopted to measure $\cFFP(x)$  kept fixed. Lines correspond
to combined fits: we model data with b.c. type 1 according to
$y^{(1)} = a + b^{(1)}\sqrt{\rd t}$ and data with b.c. type 2
according to $y^{(2)} = a + b^{(2)}\sqrt{\rd t}$ and then we
minimize the overall $\chi^2$--function  
\begin{equation}
\chi^2 = \sum_{k=1}^3\left\{\left[\frac{y^{(\text{b.c. type 1, simulated})}_k -
    y^{(1)}_k}{\sigma^{(1)}_k}\right]^2+\left[\frac{y^{(\text{b.c. type 2, simulated})}_k -
    y^{(2)}_k}{\sigma^{(2)}_k}\right]^2\right\}\,,
\end{equation}
with respect to $a$, $b^{(1)}$ and $b^{(2)}$. Fits are
statistically compatible with a convergence rate of $\sqrt{\rd
  t}$ for both b.c. This result agrees with the findings of
\cite{Slominski}, despite the different b.c. there considered. 

In Fig.~\ref{fig:fpdistr} (bottom), we compare the distribution of
the intra--party excess of votes obtained from the Monte Carlo simulation of
the VM at $(\omega_1,\omega_2)=(2,\!000.0,0.3)$ and $Q=6$ with the
corresponding distributions obtained by numerically integrating
eq.~(\ref{eq:ito}) at the same parameters, with $\rd t =0.1,\, 0.01,\,
0.001$ and b.c type 1 and type 2. We see that
\begin{itemize}
\item{all distributions coincide for $x\gtrsim 0.01$. Time
    discretization effects are negligible in this regime, where
    $\bar\phi(t)$ has larger components on average. Clique MFT
    matches very well the VM; }
\item{by contrast the distributions obtained from the numerical integration of
  eq.~(\ref{eq:ito})} spread out for $x\lesssim 0.01$. The lesser
$x$, the closer $\bar\phi(t)$ lies to
$\partial\cT_Q(\omega_1^{-1})$ on average. In practice, the regime of small 
$x$ corresponds to probing frequently $\partial \cT_Q(\omega_1^{-1})$, where
boundary reflections become important;
\item{$\cFFP(x)$ converges from below with b.c. type 1 as $\rd
    t\to 0$ and from above with b.c. type 2. We have already
    observed that b.c. type 1 are repulsive, whereas b.c. type 2
    are in a sense attractive. Therefore, the former depress
    $\cFFP(x)$ at small $x$ and finite $\rd t$, while the latter enhance it;}
\item{$\cFFP(x)$ has nothing to do with $\cFVM(x)$ for $x\lesssim
    0.01$. In sect.~2 we have observed that close to the lower
    bound $\omega_1^{-1}$, the variable $\phi_k$ is
    quantized and consequently $\cFVM(x)$ is a mixture of Dirac delta
    distributions. Clique MFT assumes $\phi_k$ to be a continuous
    variable, thus its probability mass is spread continuously
    across the whole support
    $[\omega_1^{-1},Q(1-\omega_1^{-1})]$. Anyway, at our
    simulation parameters, the probability content of the region
    $[\omega_1^{-1},0.01]$ amounts to $\lesssim 0.01$, so it does not
    impact visibly on the shape of the distribution at larger values of $x$;}
\item{an important point which is not discussed in
    \cite{Slominski} is the dependence of the convergence rate of
    the stochastic process as $\rd t\to 0$ upon the physical
    dimension of the convex domain $\cD$. We observe a strong
    enhancement of the time discretization effects (the slopes of
    the curves in Fig.~\ref{fig:fpdistr} (top)) as $Q$
    increases. This suggests the ansatz} 
  \begin{equation}
    \E\left\| \bar\phi(t)|_{\rd t\ne 0} - \bar\phi(t)\right\| \simeq
    \gamma(Q)\sqrt{\rd t}\,,
  \end{equation}
  with $\gamma(Q)$ being a proportionality factor increasing with
  the number of candidates and depending on the norm
  considered. Larger values of $Q$ require smaller values of $\rd
  t$ to keep time discretization effects constant. This makes the
  numerical integration of eq.~(\ref{eq:ito}) in the thermodynamic
  limit highly CPU demanding.
\end{itemize}

To sum up, clique MFT describes the VM rather well with respect to
the autocorrelation times, the average values of the vote
variables and the distribution of the excess of votes in the
regime where the scaling vote variables can be approximately assumed to
be continuous. 

\section{Conclusions}

In this paper we explored the stochastic dynamics of the
multi--state voter model over a social network, where people belong to
cliques dominated by zealot candidates and contacts among different
cliques are static and random. In the thermodynamic limit, the
model is characterized by two parameters, representing
respectively the size of the cliques and the frequency of the
random contacts. We studied the equilibrium distribution of the
vote variables by means of numerical simulations and analytic
arguments based on mean field theory.  

In a sense, the model corresponds to a reinterpretation of the
{\it word of mouth} model by Fortunato and Castellano (2007)
\cite{fcscaling} in a framework where voting is compulsory. The
distribution of the intra--party excess of votes observed by us
differs significantly from the one characterizing the original
word of mouth model, while it resembles roughly that observed in
the Brazilian elections: the almost flat shapes pointing towards
small values of the scaling vote variable (\cfr Fig.~\ref{fig:xs})
look phenomenologically encouraging (in a comparison with Fig.~2
of \cite{fempanal}), yet a serious point of weakness is
represented by the steepness of the opposite tail. It is difficult
to say whether the latter is related to an excessive rigidity of
the network considered, an excessive na\"iveness of the
interaction rule or to some other neglected factor. After all, though
inspired by real elections, the model is only a crude approximation 
of reality (\eg cliques of $\omega_1=1,\!000$ agents are certainly unrealistic),  
and its interest lies in its non--trivial dynamics. 

It would be interesting to understand how the model depends upon its
parameters over a wider spectrum of values than those chosen
in this paper and to introduce additional sources of
stochasticity (\eg by making the size of the cliques
fluctuate). It would be also interesting to
simulate the model over different topologies, where the zealot
candidates hold positions which are statistically symmetric
within the network. 

A better understanding of the analytic properties of the
distribution of the excess of votes, at least in the context of the
Fokker--Planck equation, seems to us unavoidable in order to learn
how to improve the model. Some ideas are just sketched in Appendix A,
to which the interested reader is referred.

\section*{Acknowledgments}

We thank R. Filippini for her participation in the early stage of this
work. The computing resources used for our numerical study and the related
technical support have been provided by the CRESCO/ENEA\-GRID High Performance
Computing infrastructure and its staff \cite{cresco}. CRESCO ({\color{red}
C}omputational  {\color{red} RES}earch centre on {\color{red} CO}mplex
systems) is funded by ENEA and by Italian and European research programmes. 

\begin{appendices}

\section{On polynomial approximations of $\cPFP(\bar\phi)$}

Eqs.~(\ref{eq:driftone})--(\ref{eq:diffusone}) show that the
coefficient functions $A^{(i)}_\ell$ and $B^{(i)}_{\ell m}$ of the
FP equation are polynomials of respectively first and second
degree  in $\bar\phi$. Since the drift and the diffusion terms involve
respectively one and two differentiations, it makes sense to expand
$\cPFP(\bar\phi)$ in polynomial functions. The polynomial coefficients should
be such that the FP equation is fulfilled order 
by order. Could the maximum degree of the polynomials be finite? To answer
this question, it should not be forgotten that the VM collapses to
intra--clique consensus as $\omega_2\to 0$. Although we did not perform
numerical integrations of the FP equation in this limit, it is reasonable to
expect that 
\begin{equation}
\lim_{\omega_2\to 0} \cPFP(\bar\phi)  \simeq
\prod_{i=1}^Q\prod_{\ell\ne i}^{1\ldots Q}\delta(\phi^{(i)}_\ell)\,.
\end{equation}
If this is correct, the maximum degree of an approximating polynomial
should depend on the model parameters and 
rapidly increase as $\omega_2\to 0$. Developing polynomial
approximations of $\cPFP(\bar\phi)$ is a hard task, although polynomials are
the most elementary functions one can deal with. Here, we make
do with discussing some theoretical aspects of the problem and
refer the reader to a future publication for quantitative
results~\cite{Palombi1}.  

\subsection{Projecting $\cPFP(\bar\phi)$ onto a basis of Dirichlet distributions}

We have seen in sect.~3 that the marginal distributions of
$\phi^{(i)}_\ell$ can be empirically modelled by Beta
distributions with very good match. In this respect,
eqs.~(\ref{eq:betansone})--(\ref{eq:betanstwo}) are 
highly suggestive. Recall  that a
Dirichlet distribution Dir($\alpha$) of order $d+1$ with
parameters $\alpha=\{\alpha_k\}_{k=1}^{d+1}$ has probability density
\begin{align}
& \cD_\alpha(x) =
\frac{\Gamma\left(|\alpha|_1\right)}{\prod_{k=1}^{d+1}\Gamma(\alpha_k)}
\left(\prod_{k=1}^d
  x_k^{\alpha_k-1}\right)(1-|x|_1)^{\alpha_{d+1}-1}\,,\qquad x\in{\bar T}_d\,,\\[2.0ex]
& {\bar T}_d = \{y\in\dR^d_+: |y|_1\le 1\}\,.
\end{align}
It is well known that if $X\sim\text{Dir}(\alpha)$, then
$X_i\sim\text{Beta}(\alpha_i,|\alpha|_1-\alpha_i)$. Inferring
from eqs.~(\ref{eq:betansone})--(\ref{eq:betanstwo}) that the
clique vector variable $\bar\phi^{(i)}\equiv\{\phi^{(i)}_k\}_{k\ne i}^{1\ldots,
Q}\sim\text{Dir}(\alpha,\ldots,\alpha,\bar\alpha,\alpha,\ldots,\alpha)$,
with $\alpha,\bar\alpha$ set according to Table~\ref{tab:fitpars}
and $\bar\alpha$ being the $i^\text{th}$ component of the
parameter vector, could be not far from being correct:  at least the
argument suggests that Dirichlet distributions might play
a relevant r\^ole in analytic representations of $\cPFP(\bar\phi)$. On a more
sound basis, we observe that 
\begin{equation}
\text{span}\left\{\cD_\alpha:
  \alpha\in\dN^{d+1},|\alpha|_1=n+d+1\right\} =
\text{span}\left\{x_1^{\alpha_1}\cdot\ldots\cdot
  x_d^{\alpha_d}:\alpha\in\dN_0^d,|\alpha|_1\le n\right\}\,,
\end{equation}
\ie Dirichlet distributions with positive integer
parameters form a basis of the vector space of multivariate
polynomials. Thus, a representation of $\cPFP(\bar\phi)$ as a polynomial
series can be certainly given in terms of Dirichlet distributions as
\begin{align}
& \cPFP(\bar\phi) = \hskip -0.1cm
\sum_{\underline{\alpha}}
c_{\underline{\alpha}}\,\cD_{\underline{\alpha}}(\bar\phi)\,, \qquad
\cD_{\underline{\alpha}}(\bar\phi)=\cD^{(1)}_{\alpha^{(1)}}(\bar\phi^{(1)})\,\ldots\,
\cD^{(Q)}_{\alpha^{(Q)}}(\bar\phi^{(Q)})\,,
\label{eq:PFPDir}\\[0.8ex]
& \cD^{(i)}_{\alpha^{(i)}}(\bar\phi^{(i)}) \equiv
\frac{\Gamma\left(|\alpha^{(i)}|_1\right)}
{\prod_{k=1}^{Q}\Gamma(\alpha^{(i)}_k)} \left(\prod_{k\ne
    i}^{1\ldots Q}
  \left[\phi^{(i)}_k\right]^{\alpha^{(i)}_k-1}\right)\left(1-\sum_{k\ne
  i }^{1\ldots Q}\phi^{(i)}_{k}
\right)^{\alpha^{(i)}_{i}-1}\,,\\[4.0ex]
& \sum_{\underline{\alpha}}
c_{\underline{\alpha}}=1\,,
\end{align}
where the sum in eq.~(\ref{eq:PFPDir}) extends to all combinations
$\underline{\alpha}=\{\alpha^{(1)},\ldots,\alpha^{(Q)}\}\in \dN^{Q\times Q}$ of
positive integers. In order to avoid complications, we have
deliberately neglected  the simplex cut off\, $\sum_{k\ne
  i}\phi^{(i)}_k<1-\omega_1^{-1}$ in eq.~(\ref{eq:PFPDir}): we
shall assume for the time being that the support of
$\cPFP(\bar\phi)$ is $\cT_Q(0)$ instead of $\cT_Q(\omega_1^{-1})$
and come back to this point later on. From eq.~(\ref{eq:PFPDir})
it follows  
\begin{align}
\cFFP(x) & =
\int_{\cT_Q(0)}\!\!\delta(\phi_\ell-x)\,\cPFP(\bar\phi)\,\rd\bar\phi\, = 
\sum_{\underline{\alpha}} c_{\underline{\alpha}}\,
\cF_{\underline{\alpha}}(x)\,,
\end{align}
with
\begin{equation}
\cF_{\underline{\alpha}}(x) = \int_{\cT_Q(0)}\delta(\phi_\ell-x)
\,\cD^{(1)}_{\alpha^{(1)}}(\bar\phi^{(1)})\,\ldots\,
\cD^{(Q)}_{\alpha^{(Q)}}(\bar\phi^{(Q)})\,\rd\bar\phi\,.  
\label{eq:Falpha}
\end{equation}
Before embarking on a determination of the coefficients
$c_{\underline{\alpha}}$ based on the FP
equation (which is anyway beyond the scope of this appendix), we should
investigate the analytic structure of the integrals $\cF_{\underline{\alpha}}(x)$. To
this aim, we make use of a well known formal trick based on the Fourier
representation of the Dirac delta function 
\begin{equation}
\delta(x) = \frac{1}{2\pi}\int_{-\infty}^{+\infty}\rd \lambda\ \re^{\ri\lambda x}\,,
\end{equation}
and the formalism of the Laplace transform
\begin{equation}
\cL\{f\}(p) = \int_0^\infty \re^{-p t}f(t)\,\rd t\,.
\label{eq:laptrans}
\end{equation}
To start with, we observe that by virtue of eq.~(\ref{eq:phikFC})
many variables in eq.~(\ref{eq:Falpha}) can 
be integrated out. Accordingly, $\cF_{\underline{\alpha}}(x)$ boils down to
\begin{align}
\cF_{\underline{\alpha}}(x) & = \int \delta\left(1-x + \sum_{k\ne \ell}^{1\ldots
  Q}[\phi^{(k)}_\ell-\phi^{(\ell)}_k]\right)\nonumber\\[0.0ex]
& \times \left\{\prod_{k\ne\ell}^{1\ldots Q}
  \beta_{\alpha^{(k)}_\ell,|\alpha^{(k)}|_1-\alpha^{(k)}_\ell}(\phi^{(k)}_\ell)\right\} 
\cD^{(\ell)}_{\alpha^{(\ell)}}(\bar\phi^{(\ell)})\prod_{k\ne\ell}^{1\ldots
  Q}\rd\phi^{(\ell)}_k\rd\phi^{(k)}_\ell\,. 
\label{eq:Fdistrib}
\end{align}
We process separately the three terms under the integral sign. First, we have
\begin{equation}
\delta\left(1-x+\sum_{k\ne\ell}^{1\ldots Q}[\phi^{(k)}_\ell -
  \phi^{(\ell)}_k]\right) = \frac{1}{2\pi}\int_{\infty}^{\infty}\rd q
\,\re^{\ri q\left\{1-x+\sum_{k\ne\ell}\left[\phi^{(k)}_\ell - \phi^{(\ell)}_k\right]\right\}}\,,
\label{eq:deltafour}
\end{equation}
We then represent $\cD^{(\ell)}_{\alpha^{(\ell)}}(\bar\phi^{(\ell)})$ in the equivalent form
\begin{align}
\cD^{(\ell)}_{\alpha^{(\ell)}}(\bar\phi^{(\ell)}) & =
\frac{\Gamma\left(|\alpha^{(\ell)}|_1\right)} 
{\prod_{k=1}^{Q}\Gamma(\alpha^{(\ell)}_k)} \left(\prod_{k\ne
    \ell}^{1\ldots Q}
  \left[\phi^{(\ell)}_k\right]^{\alpha^{(\ell)}_k-1}\right)\int_0^\infty\,
\left[\phi^{(\ell)}_i\right]^{\alpha^{(\ell)}_{\ell}-1}
\delta\left(1-\sum_{k=1}^Q\phi^{(\ell)}_{k} 
\right)\,\rd\phi^{(\ell)}_\ell\nonumber\\[1.0ex]
& =  \frac{\Gamma\left(|\alpha^{(\ell)}|_1\right)} 
{\prod_{k\ne \ell}^{1\ldots Q}\Gamma(\alpha^{(\ell)}_k)} \left(\prod_{k\ne
    \ell}^{1\ldots Q}
  \left[\phi^{(\ell)}_k\right]^{\alpha^{(\ell)}_k-1}\right)
\frac{1}{2\pi \ri}\int_{-\ri\infty}^{\ri\infty} 
\,\frac{\re^{p_0\left(1-\sum_{k\ne \ell}^{1\ldots
        Q}\phi^{(\ell)}_k\right)}}{p_0^{\alpha^{(\ell)}_\ell}}\,\rd
p_0\,.
\label{eq:lapdir}
\end{align}
In particular, the second line of eq.~(\ref{eq:lapdir}) follows
from {\it i}) a rotation of the Fourier integral representing the Dirac
delta to the imaginary axis, {\it ii}) a subsequent exchange of integrals
and {\it iii)} the Laplace transform of monomials. Similarly, we represent the marginal 
distribution 
$\beta_{\alpha^{(k)}_\ell,|\alpha^{(k)}|_1-\alpha^{(k)}_\ell}(\phi^{(k)}_\ell)$
as 
\begin{equation}
\beta_{\alpha^{(k)}_\ell,|\alpha^{(k)}|_1-\alpha^{(k)}_\ell}(\phi^{(k)}_\ell)
= \frac{\Gamma(|\alpha^{(k)}|_1)}{\Gamma(\alpha^{(k)}_\ell)}
\left(\phi^{(k)}_\ell\right)^{\alpha^{(k)}_\ell-1}\frac{1}{2\pi \ri}
\int_{-\ri\infty}^{+\ri\infty}\rd p\,
\frac{\re^{p\left(1-\phi^{(k)}_\ell\right)}}
{p^{|\alpha^{(k)}|_1-\alpha^{(k)}_\ell}}\,.  
\label{eq:lapbeta}
\end{equation}
In particular, from eq.~(\ref{eq:lapbeta}) it follows
\begin{align}
\prod_{k\ne\ell}^{1\ldots
  Q}\beta_{\alpha^{(k)}_\ell,|\alpha^{(k)}|_1-\alpha^{(k)}_\ell}
(\phi^{(k)}_\ell) & = \prod_{k\ne\ell}^{1\ldots
  Q}\left[\frac{\Gamma(|\alpha^{(k)}|_1)}
  {\Gamma(\alpha^{(k)}_\ell)}
  \left(\phi^{(k)}_\ell\right)^{\alpha^{(k)}_\ell-1}\right]   
\nonumber\\[1.0ex]
& \cdot \frac{1}{(2\pi \ri)^{Q-1}}\int_{-\ri\infty}^{+\ri\infty}\prod_{k\ne
  \ell}^{1\ldots Q} \left[\frac{\rd
  p_k}{p_k^{|\alpha^{(k)}|_1-\alpha^{(k)}_\ell}}\right]\re^{\sum_{k\ne\ell}
  p_k\left[1-\phi^{(k)}_\ell\right]}\,.
\label{eq:betaprod}
\end{align}
The next step is to insert eqs.~(\ref{eq:deltafour}),
(\ref{eq:lapdir}) and (\ref{eq:betaprod}) into
eq.~(\ref{eq:Fdistrib}). Internal integrals over the
$\phi$'s read
\begin{align}
& \left\{\prod_{k\ne\ell}^{1\ldots
    Q}\int_0^\infty\rd\phi^{(\ell)}_k\,\left[\phi^{(\ell)}_k\right]^{\alpha^{(\ell)}_k-1}
  \re^{-(\ri q+p_0)\phi^{(\ell)}_k}\right\}
\left\{\prod_{k\ne\ell}^{1\ldots Q}\int_0^\infty\rd\phi^{(k)}_\ell\,\left[\phi^{(k)}_\ell\right]^{\alpha^{(k)}_\ell-1}
\re^{-(p_k-\ri q)\phi^{(k)}_\ell}\right\}\nonumber\\[0.0ex] 
& = \prod_{k\ne\ell}^{1\ldots
Q} \left[\frac{\Gamma(\alpha^{(\ell)}_k)}{(p_0+\ri q)^{\alpha^{(\ell)}_k}}\right]\prod_{k\ne\ell}^{1\ldots
Q}\left[\frac{\Gamma(\alpha^{(k)}_\ell)}{(p_k-\ri q)^{\alpha^{(k)}_\ell}}\right]\,.
\end{align}
Thus, we have
\begin{align}
\cF_{\underline{\alpha}}(x) & =
\frac{\prod_{k=1}^Q\Gamma(|\alpha^{(k)}|_1)}{2\pi}
\int_{-\infty}^{+\infty} \rd q\, \re^{\ri q(1-x)}\nonumber\\[0.0ex]
& \cdot\frac{1}{2\pi
  \ri}\int_{-\ri\infty}^{+\ri\infty}\,\rd p_0\,\frac{1}{p_0^{\alpha^{(\ell)}_\ell}}\prod_{k\ne
\ell}^{1\ldots
Q}\left[\frac{1}{(p_0+\ri q)^{\alpha^{(\ell)}_k}}\right]
\re^{p_0s}\biggr|_{s=1}\nonumber\\[0.0ex]
& \cdot \frac{1}{(2\pi \ri)^{Q-1}}\prod_{k\ne \ell}^{1\ldots
  Q}\int_{-\ri\infty}^{+\ri\infty}\,\rd p_k\, \frac{\re^{p_k
    s}}{p_k^{|\alpha^{(k)}|_1-\alpha^{(k)}_\ell}(p_k-\ri q)^{\alpha^{(k)}_\ell}}\biggr|_{s=1}\,.
\label{eq:Fintermedone}
\end{align}
The second and third line of eq.~(\ref{eq:Fintermedone}) are inverse Laplace
transforms of products of functions. In order to solve the integrals, one
could work out the products in partial fractions or make use of the Laplace
convolution theorem, while recalling that 
\begin{align}
& \frac{1}{p_0^{\alpha^{(\ell)}_\ell}} =
\cL\left\{f(s)=\frac{s^{\alpha^{(\ell)}_\ell-1}}{\Gamma(\alpha^{(\ell)}_\ell)}\right\}(p_0)\,,\\
& \frac{1}{(p_0+iq)^{\sum_{k\ne l}\alpha^{(\ell)}_k}} =
\cL\left\{f(s)=\frac{s^{\sum_{k\ne \ell}\alpha^{(\ell)}_k-1}\re^{-iqs}}{\Gamma\left(\sum_{k\ne\ell}\alpha^{(\ell)}_k\right)}\right\}(p_0)\,,\\[0.0ex]
\end{align}
\begin{align}
& \frac{1}{p_k^{|\alpha^{(k)}|_1-\alpha^{(k)}_\ell}} =
\cL\left\{f(s)=\frac{s^{|\alpha^{(k)}|_1-\alpha^{(k)}_\ell-1}}{\Gamma(|\alpha^{(k)}|_1-\alpha^{(k)}_\ell)}\right\}(p_k)\,,\\
& \frac{1}{(p_k-iq)^{\alpha^{(k)}_\ell}} = \cL\left\{f(s)=\frac{s^{\alpha^{(k)}_\ell-1}\re^{iqs}}{\Gamma\left(\alpha^{(k)}_\ell\right)}\right\}(p_k)\,.
\end{align}
From the Laplace convolution theorem, it follows
\begin{align}
& \frac{1}{2\pi \ri}\int_{-\ri\infty}^{+\ri\infty}\rd
p_0\,\frac{\re^{p_0s}}{p_0^{\alpha^{(\ell)}_\ell}(p_0+\ri q)^{\sum_{k\ne\ell}\alpha^{(\ell)}_k}}\biggr|_{s=1} \nonumber\\[0.5ex]
& =\frac{1}{\Gamma(\alpha^{(\ell)}_\ell)\Gamma(\sum_{k\ne\ell}\alpha^{(\ell)}_k)}\int_0^1\rd\sigma\,
\sigma^{\sum_{k\ne\ell}\alpha^{(\ell)}_k-1}(1-\sigma)^{\alpha^{(\ell)}_\ell-1}\re^{-\ri
  q\sigma}\nonumber\\[0.5ex]
& =
\frac{1}{\Gamma\left(|\alpha^{(\ell)}|_1\right)}M\left(\sum_{k\ne\ell}\alpha^{(\ell)}_k,|\alpha^{(\ell)}|_1,-\ri
  q\right)\,,
\label{eq:lapone}
\end{align}
with $M(a,b,z)={}_1F_1(a;b;z)$ being the Kummer function. Analogously, we have
\begin{align}
& \frac{1}{(2\pi \ri)^{Q-1}}\prod_{k\ne\ell}^{1\ldots
  Q}\int_{-\ri\infty}^{+\ri\infty}\rd p_k\,
\frac{\re^{p_ks}}{p_k^{|\alpha^{(k)}|_1-\alpha^{(k)}_\ell}(p_k-\ri
  q)^{\alpha^{(\kappa)}_\ell}}\biggr|_{s=1}\nonumber\\[0.5ex]
& =\prod_{k\ne\ell}^{1\ldots
  Q}\left\{\frac{1}{\Gamma(\alpha^{(k)}_\ell)\Gamma\left(|\alpha^{(k)}|_1-\alpha^{(k)}_\ell\right)}\int_0^1\rd\sigma\,\sigma^{\alpha^{(k)}_\ell-1}(1-\sigma)^{|\alpha^{(k)}|_1-\alpha^{(k)}_\ell-1}\,\re^{\ri
    q\sigma}\right\}\nonumber\\[0.5ex]
& = \prod_{k\ne\ell}^{1\ldots
  Q}\left\{\frac{1}{\Gamma\left(|\alpha^{(k)}|_1\right)}M\left(\alpha^{(k)}_\ell,|\alpha^{(k)}|_1,\ri
    q\right)\right\}\,.
\label{eq:laptwo}
\end{align}
Inserting the right--hand side of eqs.~(\ref{eq:lapone})--(\ref{eq:laptwo})
back into eq.~(\ref{eq:Fintermedone}) and using the Kummer identity (\cfr
eq.~(13.1.27) of \cite{abramowitz})
\begin{equation}
M\left(\sum_{k\ne\ell}\alpha^{(\ell)}_k,|\alpha^{(\ell)}|_1,-\ri q\right)
= \re^{-\ri q}M\left(\alpha^{(\ell)}_\ell,|\alpha^{(\ell)}|_1,\ri q\right)\,,
\end{equation}
yields
\begin{equation}
\cF_{\underline{\alpha}}(x) = \frac{1}{2\pi}\int_{-\infty}^{+\infty}\rd
q\,\re^{-\ri
  qx}\,\phantom{}\prod_{k=1}^QM\left(\alpha^{(k)}_\ell,|\alpha^{(k)}|_1,\ri
  q\right)\,,
\label{eq:Fkk}
\end{equation}
or equivalently (by setting $p=-\ri q$), 
\begin{equation}
\cF_{\underline{\alpha}}(x) = \frac{1}{2\pi \ri}\int_{-\ri\infty}^{+\ri\infty}\rd
p\ \re^{px}\,\phantom{}\prod_{k=1}^Q M\left(\alpha^{(k)}_\ell,|\alpha^{(k)}|_1,-p\right)\,.
\label{eq:finalF}
\end{equation}
Thus, we see that the contributions to $\cFFP(x)$ arising from a polynomial expansion of
$\cPFP(\bar\phi)$ based on Dirichlet distributions are higher transcendental
functions. Alternatively, by using once again the Laplace convolution
theorem, we can express $\cF_{\underline{\alpha}}(x)$ as a nested integral
\begin{align}
\cF_{\underline{\alpha}}(x) & =
\int_0^x\rd\tau_1\int_0^{\tau_1}\rd\tau_2\ldots\int_0^{\tau_{Q-2}}\rd\tau_{Q-1}\,W(\alpha^{(\ell)}_\ell,|\alpha^{(\ell)}|_1,x-\tau_1)\nonumber\\[1.0ex]
& \hskip 2.0cm \cdot
W(\alpha^{(1)}_\ell,|\alpha^{(1)}|_1,\tau_1-\tau_2)\ldots W(\alpha^{(Q)}_\ell,|\alpha^{(Q)}|_1,\tau_{Q-1})\,,
\end{align}
with 
\begin{equation}
W(a,b,t) =
\frac{\Gamma(b)}{\Gamma(a)\Gamma(b-a)}t^{a-1}(1-t)^{b-a-1}\theta(1-t) = \beta_{a,b-a}(t)\,\theta(1-t)\,,
\end{equation}
and $\theta(\cdot)$ being the Heaviside step function. Eq.~(\ref{eq:Fkk}) can
be computed numerically once $\underline{\alpha}$ is assigned. As an example,
we consider the subset of indices
\begin{equation}
\underline{\alpha} = \underline{\alpha}(\kappa,\bar\kappa)\,\equiv\, \left\{\begin{array}{ll}
\alpha^{(i)}_j = \kappa\,, & \quad i\ne j \text{\ \, and
  \ }i,j=1,\ldots,Q\,,\\[1.0ex] 
\alpha^{(i)}_i=\bar\kappa\,, & \quad  i=1,\ldots, Q\,,
\end{array}\right.
\end{equation}
corresponding to the contributions
\begin{equation}
\cF_{\underline{\alpha}(\kappa,\bar\kappa)}(x) =
\frac{1}{2\pi}\int_{-\infty}^{+\infty}\rd q\, \re^{-\ri
  qx}\,\left[M(\kappa,\bar\kappa+(Q-1)\kappa,\ri q)\right]^{Q-1}\,M(\bar\kappa,\bar\kappa+(Q-1)\kappa,\ri
q)\,.
\end{equation}
\begin{figure}[t!]
    \centering
    \includegraphics[width=0.8\textwidth]{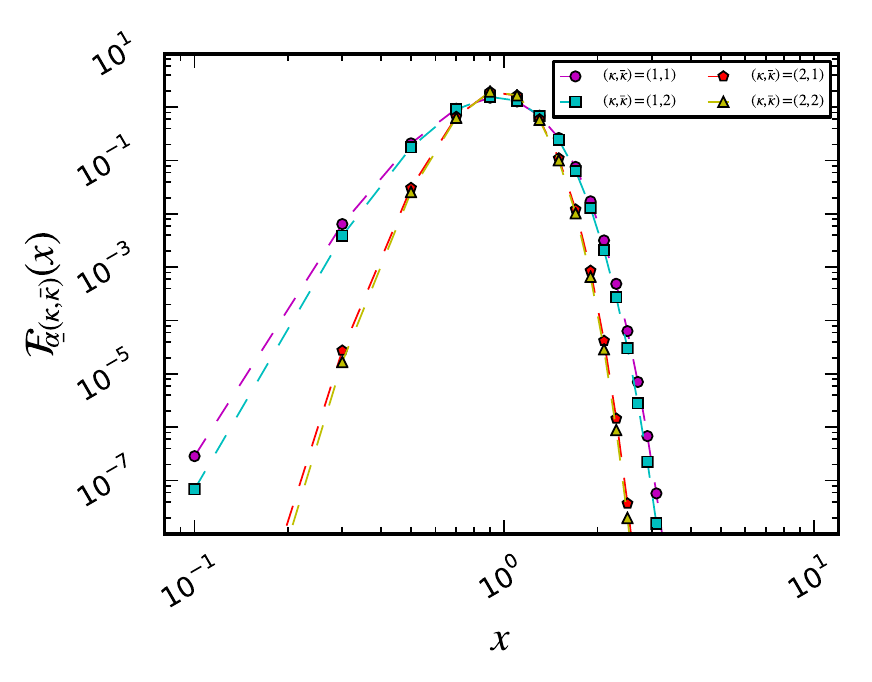}
    \vskip -0.2cm
  \caption{ \footnotesize Some contributions to $\cFFP(x)$ arising from a polynomial
    expansion of $\cPFP(\bar\phi)$ based on Dirichlet distributions.}
  \label{fig:Fkk} 
\end{figure}
In Fig.~(\ref{fig:Fkk}) we show
$\cF_{\underline{\alpha}(\kappa,\bar\kappa)}$ for some choices of
$(\kappa,\bar\kappa)$. We see that all curves lie below $\cFFP(x)$ by orders
of magnitude along the tails, thus suggesting that contributions $\cF_{\underline{\alpha}}(x)$, where the
indices $\alpha^{(i)}_j$ are not the same for $i\ne j$, cannot be neglected.  

\subsection{Cutting off the simplex}

We finally explore how the Dirichlet distribution changes upon cutting off its
defining domain. Specifically, we consider
\begin{equation}
\cD_\alpha(x) = Z_\alpha(\omega_1^{-1}) \left(\prod_{k=1}^d
  x_k^{\alpha_k-1}\right)(1-|x|_1)^{\alpha_{d+1}-1}\,,
\end{equation}
as a distribution with support in
\begin{equation}
{\bar T}_d(\omega_1^{-1}) = \{y\in\dR^d_+: |y|_1\le 1-\omega_1^{-1}\}\,.
\end{equation}
The normalization constant $Z_\alpha(\omega_1^{-1})$ can be calculated analytically according
to the same technique used for $\cF_{\underline{\alpha}}(x)$ by imposing that the
integral of $\cD_\alpha(x)$ over ${\bar T}_d(\omega_1^{-1})$ equals one. A
little algebra yields 
\begin{equation}
[Z_\alpha(\omega_1^{-1})]^{-1} = \frac{1}{2\pi \ri}\int_{-\ri\infty}^{+\ri\infty}\rd \lambda\ \re^{\lambda t}\ \left(\prod_{i=1}^{d} \int_0^{+\infty}\rd x_i\ x_i^{\alpha_i-1} \re^{-\lambda x_i}\right)\int_{\omega_1^{-1}}^{+\infty}\rd x_{d+1}\ x_{d+1}^{\alpha_{d+1}-1} \re^{-\lambda x_{d+1}}\biggr|_{t=1}\,,
\end{equation}
We notice that
\begin{equation}
\int_{\omega_1^{-1}}^{+\infty}\rd z\ z^{\alpha_{d+1}-1} \re^{-\lambda z} = \frac{\Gamma(\alpha_{d+1},\omega_1^{-1}\lambda)}{\lambda^{\alpha_{d+1}}}\,,
\end{equation}
with $\Gamma(\alpha_{d+1},\omega_1^{-1}\lambda)$ denoting the lower incomplete
Gamma function. Hence, it follows
\begin{equation}
[Z_{\alpha}(\omega_1^{-1})]^{-1} = \dfrac{\prod_{i=1}^d\Gamma(\alpha_i)}{2\pi \ri}\int_{-\ri\infty}^{+\ri\infty}\rd \lambda\ \re^{\lambda t}\ \frac{1}{\lambda^{\psi}}\frac{\Gamma(\alpha_{d+1},\omega_1^{-1}\lambda)}{\lambda^{\alpha_{d+1}}}\biggr|_{t=1}\,,
\qquad \psi = \sum_{i=1}^d \alpha_i\,.
\end{equation}
Once more, the latter integral is the Laplace antitransform of a product of functions. According to the rules of the Laplace transform, it amounts to the convolution
of the antitransform of the single functions. Therefore, we have
\begin{equation}
[Z_\alpha(\omega_1^{-1})]^{-1} =
\dfrac{\prod_{i=1}^{d+1}\Gamma(\alpha_i)}{\Gamma(|\alpha|_1)}I_{\psi,\alpha_{d+1}}(1-\omega_1^{-1})
= [Z_{{\alpha}}(0)]^{-1} I_{\psi,\alpha_{d+1}}(1-\omega_1^{-1})\,,
\end{equation}
where
\begin{equation}
I_{a,b}(z) = \int_0^z \rd \tau\ \beta_{a,b}(\tau)\,,
\end{equation}
denotes the regularized incomplete beta function. Since this has an asymptotic expansion at $z=1$ 
\begin{equation} 
I_{a,b}(z) = 1 - \frac{(1-z)^b z^a}{bB_{a,b}}\sum_{k=0}^\infty \frac{(-1)^k (a+b)_k (z-1)^k}{(b+1)_k}\,,\qquad b\ne -1\,,
\label{eq:Iasymp}
\end{equation}
with $B_{a,b}=\Gamma(a)\Gamma(b)/\Gamma(a+b)$, we conclude that
\begin{equation}
[Z_{{\alpha}}(\omega_1^{-1})]^{-1} = [Z_{{\alpha}}(0)]^{-1} + \rO\left[(\omega_1^{-1})^{\alpha_{d+1}}\right]\,.
\end{equation}
Thus, we see that depending on $\alpha_{d+1}$, the impact of cutting off the
integration domain (the Cartesian product of simplices), at the boundary on the Dirichlet
integrals could be very small.  

\end{appendices}

\bibliographystyle{unsrt}   
\bibliography{main}       

\begin{thebibliography}{10}

\bibitem{fcscaling}
S.~Fortunato and C.~Castellano.
\newblock {Scaling and Universality in Proportional Elections}.
\newblock {\em Physical Review Letters}, 99(13):138701, 2007.

\bibitem{fempanal}
A.~Chatterjee, M.~Mitrovi{\'c}, and S.~Fortunato.
\newblock {Universality in voting behavior: an empirical analysis}.
\newblock {\em Scientific Reports}, 3, 2013.

\bibitem{CostaFilho}
R.~N. Costa~Filho, M.~P. Almeida, J.~S. Andrade, and J.~E. Moreira.
\newblock {Scaling behavior in a proportional voting process}.
\newblock {\em Physical Review E}, 60(1):1067--1068, 1999.

\bibitem{Clifford1973}
P.~Clifford and A.~Sudbury.
\newblock A model for spatial conflict.
\newblock {\em Biometrika}, 60(3):581--588, 1973.

\bibitem{Holley1975}
R.~Holley and T.~M. Liggett.
\newblock Ergodic theorems for weakly interacting infinite systems and the
  voter model.
\newblock {\em The Annals of Probability}, 3(4):643--663, 1975.

\bibitem{Bohme}
G.~A. B\"ohme and T.~Gross.
\newblock Fragmentation transitions in multistate voter models.
\newblock {\em Physical Review E}, 85:066117, 2012.

\bibitem{Hubbell}
S.~P. Hubbell.
\newblock {\em {The Unified Neutral Theory of Biodiversity and Biogeography
  (MPB-32) (Monographs in Population Biology)}}.
\newblock Princeton University Press, 2001.

\bibitem{McKane}
A.~J. McKane, D.~Alonso, and R.~V. Sol\'e.
\newblock Analytic solution of hubbell's model of local community dynamics.
\newblock {\em Theoretical Population Biology}, 65(1):67 -- 73, 2004.

\bibitem{Pigolotti}
S.~Pigolotti, A.~Flammini, M.~Marsili, and A.~Maritan.
\newblock {Species lifetime distribution for simple models of ecologies}.
\newblock {\em Proceedings of the National Academy of Sciences USA},
  102(44):15747--51+, 2005.

\bibitem{Starnini}
M.~Starnini, A.~Baronchelli, and R.~Pastor-Satorras.
\newblock {Ordering dynamics of the multi-state voter model}.
\newblock {\em Journal of Statistical Mechanics}, P10027, 2012.

\bibitem{Mobilia2003}
M.~Mobilia.
\newblock Does a single zealot affect an infinite group of voters?
\newblock {\em Physical Review Letters}, 91:028701, 2003.

\bibitem{Acemoglu2010}
D.~Acemoglu, G.~Como, F.~Fagnani, and A.~E. Ozdaglar.
\newblock Opinion fluctuations and disagreement in social networks.
\newblock Levine's working paper archive, D.~K.~Levine, 2010.

\bibitem{Yildiz2012}
E.~Yildiz, D.~Acemoglu, A.~E. Ozdaglar, A.~Saberi, and A.~Scaglione.
\newblock Discrete opinion dynamics with stubborn agents.
\newblock {\em LIDS report 2858, to appear in ACM Transactions on Economics and
  Computation}, 2012.

\bibitem{Wu2012}
Y.~Wu and J.~Shen.
\newblock {Opinion dynamics with stubborn vertices.}
\newblock {\em Electronic Journal of Linear Algebra}, 23:790--800, 2012.

\bibitem{Xie1}
J.~{Xie}, S.~{Sreenivasan}, G.~{Korniss}, W.~{Zhang}, C.~{Lim}, and B.~K.
  {Szymanski}.
\newblock {Social consensus through the influence of committed minorities}.
\newblock {\em Physical Review E}, 84(1):011130, 2011.

\bibitem{Xie2}
J.~Xie, J.~Emenheiser, M.~Kirby, S.~Sreenivasan, B.~K. Szymanski, and
  G.~Korniss.
\newblock Evolution of opinions on social networks in the presence of competing
  committed groups.
\newblock {\em PLoS ONE}, 7(3):e33215, 2012.

\bibitem{Singh}
P.~Singh, S.~Sreenivasan, B.~K. Szymanski, and G.~Korniss.
\newblock Accelerating consensus on coevolving networks: The effect of
  committed individuals.
\newblock {\em Physical Review E}, 85:046104, 2012.

\bibitem{mobilia2013}
M.~Mobilia.
\newblock {Commitment Versus Persuasion in the Three-Party Constrained Voter
  Model}.
\newblock {\em Journal of Statistical Physics}, 151:69--91, 2013.

\bibitem{wattz1998}
D.~J. Watts and S.~H. Strogatz.
\newblock {Collective dynamics of ``small--world'' networks.}
\newblock {\em Nature}, 393(6684):409--10, 1998.

\bibitem{Gardiner}
C.~W. Gardiner.
\newblock {\em Handbook of Stochastic Methods}.
\newblock Springer Series in Synergetics. Springer, 1994.

\bibitem{ICWSM09154}
M.~Bastian, S.~Heymann, and M.~Jacomy.
\newblock Gephi: An open source software for exploring and manipulating
  networks, 2009.

\bibitem{Mobilia2007}
M.~Mobilia, A.~Petersen, and S.~Redner.
\newblock {On the role of zealotry in the voter model}.
\newblock {\em Journal of Statistical Mechanics: Theory and Experiment},
  2007(08):P08029+, 2007.

\bibitem{Maruyama}
G.~Maruyama.
\newblock {Continuous Markov processes and stochastic equations}.
\newblock {\em Rendiconti del Circolo Matematico di Palermo}, 4(1):48--90,
  1955.

\bibitem{Slominski}
L.~S{\l}omi\'nski.
\newblock On approximation of solutions of multidimensional sde's with
  reflecting boundary conditions.
\newblock {\em Stochastic Processes and their Applications}, 50(2):197--219,
  1994.

\bibitem{cresco}
see {\tt http://www.cresco.enea.it/english} for information.

\bibitem{Palombi1}
F.~Palombi and S.~Toti.
\newblock {Work} in progress.

\bibitem{abramowitz}
M.~Abramowitz and I.~A. Stegun.
\newblock {\em Handbook of Mathematical Functions with Formulas, Graphs, and
  Mathematical Tables}.
\newblock Dover Publications, New York, 1964.

\end{thebibliography}

\end{document}